\pdfoutput=1
\documentclass[12pt,a4paper]{article}

\usepackage{ifthen} 
\usepackage{verbatim}
\usepackage{overpic}

\newboolean{pdflatex}
\setboolean{pdflatex}{true} 

\newboolean{articletitles}
\setboolean{articletitles}{true} 

\newboolean{uprightparticles}
\setboolean{uprightparticles}{false} 

\def\paperauthors{LHCb collaboration} 
\def\paperasciititle{Measurement of Upsilon production cross-section in pp collisions at sqrt(s)=5 TeV} 
\def\papertitle{Measurement of $\Upsilonres$ production in $pp$ collisions at $\sqs=5\tev$} 
\def\paperkeywords{{High Energy Physics}, {LHCb}} 
\def\papercopyright{\the\year\ CERN for the benefit of the LHCb collaboration} 
\def\paperlicence{CC BY 4.0 licence}
\def\paperlicenceurl{https://creativecommons.org/licenses/by/4.0/}

\usepackage[top=1in, bottom=1.25in, left=1in, right=1in]{geometry}

\usepackage{microtype}
\usepackage{lineno}  
\usepackage{xspace} 
\usepackage{caption} 

\usepackage{graphicx}  
\usepackage{color}
\usepackage{colortbl}
\graphicspath{{./figs/}} 

\usepackage{amsmath} 
\usepackage{amssymb}
\usepackage{amsfonts}
\usepackage{upgreek} 

\newcommand*\patchAmsMathEnvironmentForLineno[1]{%
\expandafter\let\csname old#1\expandafter\endcsname\csname #1\endcsname
\expandafter\let\csname oldend#1\expandafter\endcsname\csname
end#1\endcsname
 \renewenvironment{#1}%
   {\linenomath\csname old#1\endcsname}%
   {\csname oldend#1\endcsname\endlinenomath}%
}
\newcommand*\patchBothAmsMathEnvironmentsForLineno[1]{%
  \patchAmsMathEnvironmentForLineno{#1}%
  \patchAmsMathEnvironmentForLineno{#1*}%
}
\AtBeginDocument{%
\patchBothAmsMathEnvironmentsForLineno{equation}%
\patchBothAmsMathEnvironmentsForLineno{align}%
\patchBothAmsMathEnvironmentsForLineno{flalign}%
\patchBothAmsMathEnvironmentsForLineno{alignat}%
\patchBothAmsMathEnvironmentsForLineno{gather}%
\patchBothAmsMathEnvironmentsForLineno{multline}%
\patchBothAmsMathEnvironmentsForLineno{eqnarray}%
}

\usepackage{hyperxmp}

\usepackage[pdftex,
            pdfauthor={\paperauthors},
            pdftitle={\paperasciititle},
            pdfkeywords={\paperkeywords},
            pdfcopyright={Copyright (C) \papercopyright},
            pdflicenseurl={\paperlicenceurl}]{hyperref}

\usepackage[colorinlistoftodos,textsize=scriptsize]{todonotes}

\usepackage[bottom,flushmargin,hang,multiple]{footmisc}

\usepackage[all]{hypcap} 

\usepackage{xspace} 
\usepackage{upgreek}

\def\lhcb   {\mbox{LHCb}\xspace}

\def\MagUp {\mbox{\em Mag\kern -0.05em Up}\xspace}

\ifthenelse{\boolean{uprightparticles}}%
{

 \def\Pmu         {\ensuremath{\upmu}\xspace}

 \def\Ppsi        {\ensuremath{\uppsi}\xspace}

 \def\PDelta      {\ensuremath{\Delta}\xspace}                 
 \def\PXi         {\ensuremath{\Xi}\xspace}                 
 \def\PLambda     {\ensuremath{\Lambda}\xspace}                 
 \def\PSigma      {\ensuremath{\Sigma}\xspace}                 
 \def\POmega      {\ensuremath{\Omega}\xspace}                 
 \def\PUpsilon    {\ensuremath{\Upsilon}\xspace}
 \let\oldPi\Pi
 \def\PPi         {\ensuremath{\oldPi}\xspace}

 \def\PB      {\ensuremath{\mathrm{B}}\xspace}                 
                  
 \def\PD      {\ensuremath{\mathrm{D}}\xspace}

 \def\PJ      {\ensuremath{\mathrm{J}}\xspace}                 
 \def\PK      {\ensuremath{\mathrm{K}}\xspace}

 \def\Pb      {\ensuremath{\mathrm{b}}\xspace}                 
 \def\Pc      {\ensuremath{\mathrm{c}}\xspace}

 \def\Pi      {\ensuremath{\mathrm{i}}\xspace}

 \def\Ps      {\ensuremath{\mathrm{s}}\xspace}

 \def\thebaroffset{0.0em}
}
{

 \def\Pmu         {\ensuremath{\mu}\xspace}

 \def\Ppsi        {\ensuremath{\psi}\xspace}                 
                  
 \mathchardef\PDelta="7101
 \mathchardef\PXi="7104
 \mathchardef\PLambda="7103
 \mathchardef\PSigma="7106
 \mathchardef\POmega="710A
 \mathchardef\PUpsilon="7107
 \mathchardef\PPi="7105
                  
 \def\PB      {\ensuremath{B}\xspace}                 
                  
 \def\PD      {\ensuremath{D}\xspace}

 \def\PJ      {\ensuremath{J}\xspace}                 
 \def\PK      {\ensuremath{K}\xspace}

 \def\Pb      {\ensuremath{b}\xspace}                 
 \def\Pc      {\ensuremath{c}\xspace}

 \def\Pi      {\ensuremath{i}\xspace}

 \def\Ps      {\ensuremath{s}\xspace}

 \def\thebaroffset{0.18em}
}
\newcommand{\offsetoverline}[2][\thebaroffset]{\kern #1\overline{\kern -#1 #2}}%

\makeatletter
\ifcase \@ptsize \relax
  \newcommand{\miniscule}{\@setfontsize\miniscule{4}{5}}
\or
  \newcommand{\miniscule}{\@setfontsize\miniscule{5}{6}}
\or
  \newcommand{\miniscule}{\@setfontsize\miniscule{5}{6}}
\fi
\makeatother

\DeclareRobustCommand{\optbar}[1]{\shortstack{{\miniscule (\rule[.5ex]{1.25em}{.18mm})}
  \\ [-.7ex] $#1$}}

\def\mup        {{\ensuremath{\Pmu^+}}\xspace}
\def\mun        {{\ensuremath{\Pmu^-}}\xspace} 

\def\mumu       {{\ensuremath{\Pmu^+\Pmu^-}}\xspace}

\def\squark    {{\ensuremath{\Ps}}\xspace}

\def\cquark    {{\ensuremath{\Pc}}\xspace}

\def\bquark    {{\ensuremath{\Pb}}\xspace}

\def\KorKbar {\kern \thebaroffset\optbar{\kern -\thebaroffset \PK}{}\xspace}

\def\D       {{\ensuremath{\PD}}\xspace}

\def\DorDbar {\kern \thebaroffset\optbar{\kern -\thebaroffset \PD}\xspace}

\def\Dp      {{\ensuremath{\D^+}}\xspace}
\def\Dm      {{\ensuremath{\D^-}}\xspace}

\def\DpDm    {\ensuremath{\Dp {\kern -0.16em \Dm}}\xspace}

\def\B       {{\ensuremath{\PB}}\xspace}

\def\BorBbar {\kern \thebaroffset\optbar{\kern -\thebaroffset \PB}\xspace}

\def\Bd      {{\ensuremath{\B^0}}\xspace}

\def\BdorBdbar {\kern \thebaroffset\optbar{\kern -\thebaroffset \Bd}\xspace}

\def\Bs      {{\ensuremath{\B^0_\squark}}\xspace}

\def\BsorBsbar {\kern \thebaroffset\optbar{\kern -\thebaroffset \Bs}\xspace}

\def\jpsi     {{\ensuremath{{\PJ\mskip -3mu/\mskip -2mu\Ppsi}}}\xspace}

\def\Upsilonres  {{\ensuremath{\PUpsilon}}\xspace}
\def\Y#1S{\ensuremath{\PUpsilon{(#1S)}}\xspace}
\def\OneS  {{\Y1S}\xspace}
\def\TwoS  {{\Y2S}\xspace}
\def\ThreeS{{\Y3S}\xspace}

\def\LorLbar     {\kern \thebaroffset\optbar{\kern -\thebaroffset \PLambda}\xspace}

\def\to                 {\ensuremath{\rightarrow}\xspace}

\newcommand{\etot}{{\ensuremath{\varepsilon_{\mathrm{ tot}}}}\xspace}

\def\AT#1     {\ensuremath{A_{\mathrm{T}}^{#1}}\xspace}           

\def\C#1      {\ensuremath{\mathcal{C}_{#1}}\xspace}                       
\def\Cp#1     {\ensuremath{\mathcal{C}_{#1}^{'}}\xspace}                    
\def\Ceff#1   {\ensuremath{\mathcal{C}_{#1}^{\mathrm{(eff)}}}\xspace}        
\def\Cpeff#1  {\ensuremath{\mathcal{C}_{#1}^{'\mathrm{(eff)}}}\xspace}       
\def\Ope#1    {\ensuremath{\mathcal{O}_{#1}}\xspace}                       
\def\Opep#1   {\ensuremath{\mathcal{O}_{#1}^{'}}\xspace}                    

\newcommand{\aunit}[1]{\ensuremath{\text{\,#1}}}       

\newcommand{\tev}{\aunit{Te\kern -0.1em V}\xspace}
\newcommand{\gev}{\aunit{Ge\kern -0.1em V}\xspace}
\newcommand{\mev}{\aunit{Me\kern -0.1em V}\xspace}
\newcommand{\kev}{\aunit{ke\kern -0.1em V}\xspace}
\newcommand{\ev}{\aunit{e\kern -0.1em V}\xspace}
 
\newcommand{\mevc}{\ensuremath{\aunit{Me\kern -0.1em V\!/}c}\xspace}
\newcommand{\gevc}{\ensuremath{\aunit{Ge\kern -0.1em V\!/}c}\xspace}
\newcommand{\mevcc}{\ensuremath{\aunit{Me\kern -0.1em V\!/}c^2}\xspace}
\newcommand{\gevcc}{\ensuremath{\aunit{Ge\kern -0.1em V\!/}c^2}\xspace}

\def\nb {\aunit{nb}\xspace}

\def\pb {\aunit{pb}\xspace}
\def\invpb {\ensuremath{\pb^{-1}}\xspace}

\newcommand{\stat}{\aunit{(stat)}\xspace}
\newcommand{\syst}{\aunit{(syst)}\xspace}

\def\deriv {\ensuremath{\mathrm{d}}}

\def\gsim{{~\raise.15em\hbox{$>$}\kern-.85em
          \lower.35em\hbox{$\sim$}~}\xspace}
\def\lsim{{~\raise.15em\hbox{$<$}\kern-.85em
          \lower.35em\hbox{$\sim$}~}\xspace}

\def\sPlot{\mbox{\em sPlot}\xspace}

\def\sqs   {\ensuremath{\protect\sqrt{s}}\xspace}
\def\sqsnn {\ensuremath{\protect\sqrt{s_{\scriptscriptstyle\text{NN}}}}\xspace}
\def\pt         {\ensuremath{p_{\mathrm{T}}}\xspace}

\def\ptot       {\ensuremath{p}\xspace}

\newcommand{\lum} {\ensuremath{\mathcal{L}}\xspace}

\def\evtgen     {\mbox{\textsc{EvtGen}}\xspace}

\def\geant      {\mbox{\textsc{Geant4}}\xspace}

\def\photos     {\mbox{\textsc{Photos}}\xspace}

\def\pythia     {\mbox{\textsc{Pythia}}\xspace}

\def\tell1  {TELL1\xspace}
\def\ukl1   {UKL1\xspace}

\newcommand{\phz}{\phantom{0}}

\newcommand{\lhcborcid}[1]{\href{https://orcid.org/#1}{\hspace*{0.1em}\raisebox{-0.45ex}{\includegraphics[width=1em]{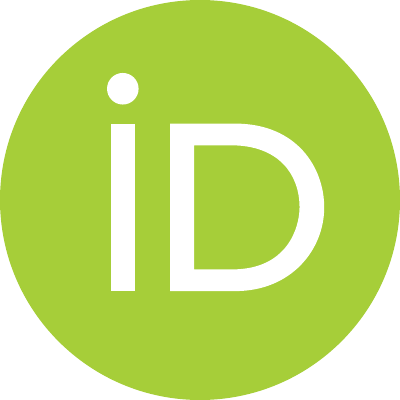}}}}

\usepackage{cite} 
\usepackage{mciteplus}

\usepackage{longtable}
\usepackage{multirow}
\usepackage{rotating}
\begin{document}

\renewcommand{\thefootnote}{\fnsymbol{footnote}}
\setcounter{footnote}{1}


\begin{titlepage}
\pagenumbering{roman}

\vspace*{-1.5cm}
\centerline{\large EUROPEAN ORGANIZATION FOR NUCLEAR RESEARCH (CERN)}
\vspace*{1.5cm}
\noindent
\begin{tabular*}{\linewidth}{lc@{\extracolsep{\fill}}r@{\extracolsep{0pt}}}
\ifthenelse{\boolean{pdflatex}}
{\vspace*{-1.5cm}\mbox{\!\!\!\includegraphics[width=.14\textwidth]{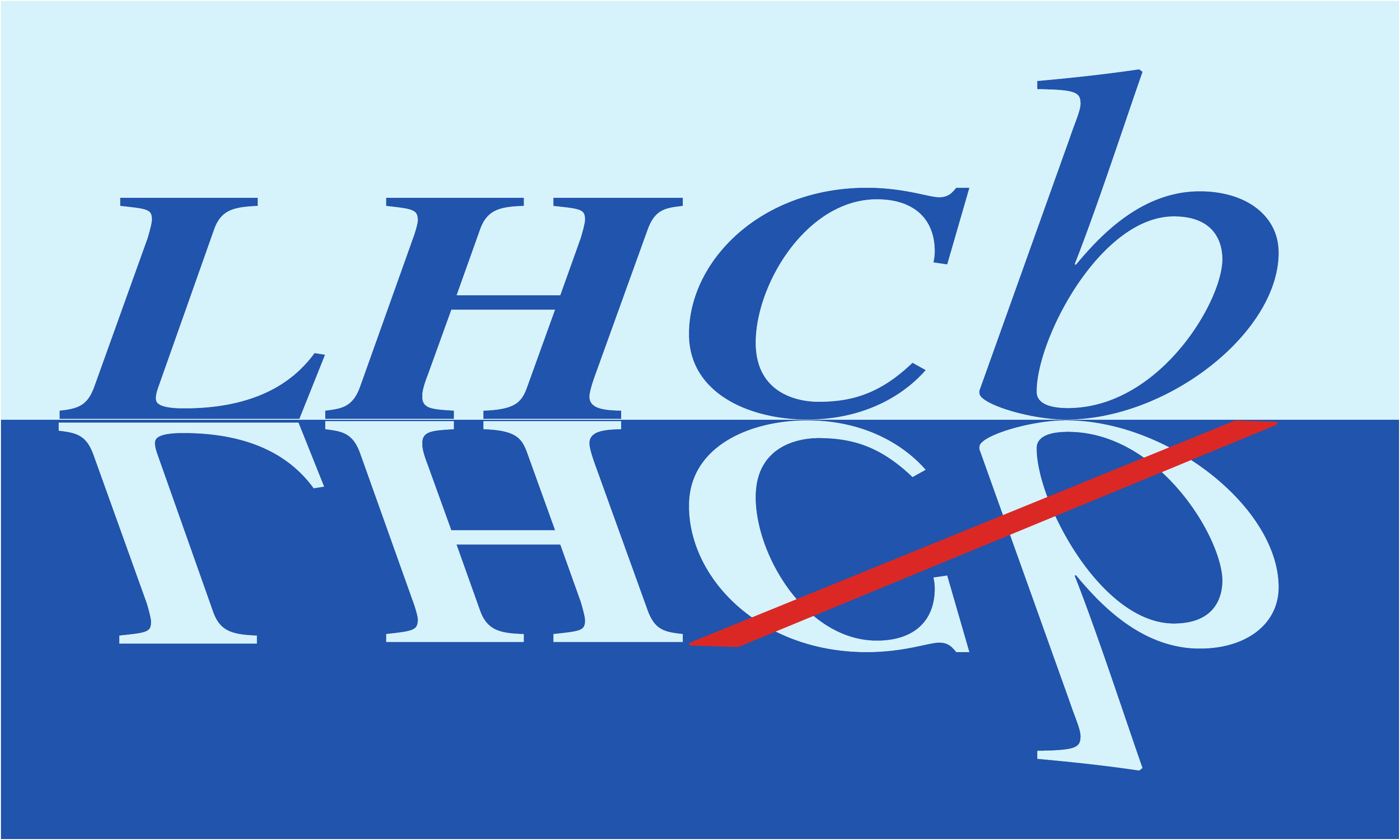}} & &}%
{\vspace*{-1.2cm}\mbox{\!\!\!\includegraphics[width=.12\textwidth]{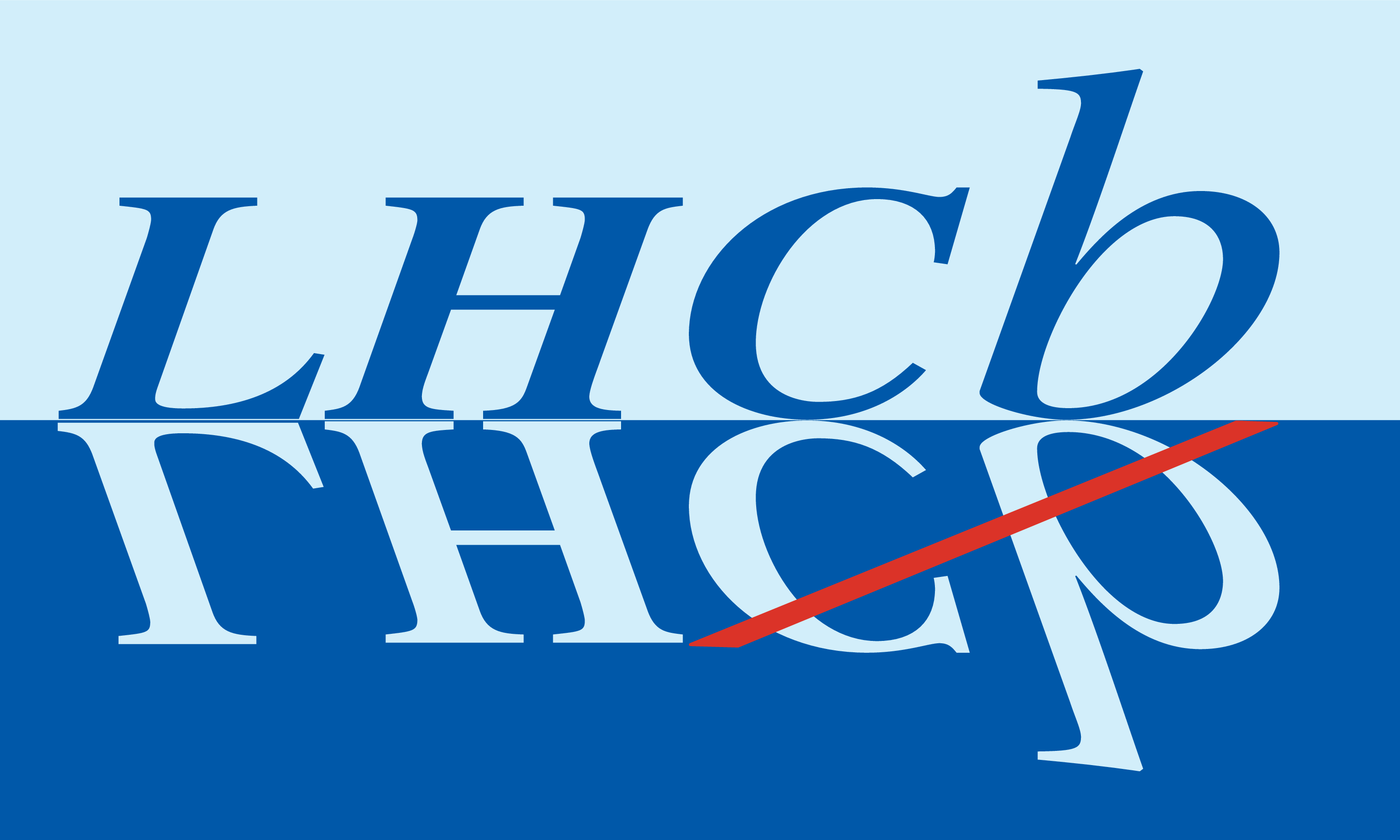}} & &}%
\\
 & & CERN-EP-2022-269 \\  
 & & LHCb-PAPER-2022-036 \\  
 & & \today \\ 
 & & \\
\end{tabular*}

\vspace*{1.0cm}

{\normalfont\bfseries\boldmath\huge
\begin{center}
  \papertitle 
\end{center}
}

\vspace*{1.0cm}

\begin{center}
\paperauthors\footnote{Authors are listed at the end of this paper.}
\end{center}

\vspace{\fill}

\begin{abstract}
  \noindent
The production cross-sections of $\Upsilonres$ mesons, namely $\OneS$, $\TwoS$ and $\ThreeS$, in $pp$ collisions at $\sqrt{s}=5 \tev$ are measured with a data sample corresponding to  an integrated luminosity of $9.13\pm{0.18} \invpb$ collected by the LHCb detector. 
The $\Upsilonres$  mesons are reconstructed in the decay mode $\Upsilonres\to\mumu$. 
Double differential cross-sections times branching fractions, as functions of the transverse momentum $\pt$ and the rapidity $y$ of the $\Upsilonres$ mesons, are measured in the range $\pt<20 \gevc$ and $2.0<y<4.5$. 
The results integrated over these $\pt$ and $y$ ranges
are
  \begin{eqnarray}
  \sigma(\OneS) \times\mathcal{B}(\emph{\OneS}\to\mumu)&=&2101\pm33\pm83\pb,\nonumber \\ 
  \sigma(\TwoS) \times\mathcal{B}(\emph{\TwoS}\to\mumu)&=&\phantom0 526\pm20\pm21 \pb,\nonumber\\ 
  \sigma(\ThreeS) \times\mathcal{B}(\emph{\ThreeS}\to\mumu)&=&\phantom0 242\pm16\pm10 \pb,\nonumber
  \end{eqnarray}
  where the first uncertainties are statistical and the second are systematic.
  The ratios of cross-sections between measurements of two different $\Upsilonres$ states and between measurements at different centre-of-mass energies are determined. 
  The nuclear modification factor of $\OneS$ at $\sqrt{s}=5 \tev$ is updated as well using the directly measured cross-section results from this analysis.
  
\end{abstract}

\vspace*{2.0cm}

\begin{center}
  Published in JHEP 07 (2023) 069 
\end{center}

\vspace{\fill}

{\footnotesize 
\centerline{\copyright~\papercopyright. \href{\paperlicenceurl}{\paperlicence}.}}
\vspace*{2mm}

\end{titlepage}


\newpage
\setcounter{page}{2}
\mbox{~}


\renewcommand{\thefootnote}{\arabic{footnote}}
\setcounter{footnote}{0}

\cleardoublepage


\pagestyle{plain} 
\setcounter{page}{1}
\pagenumbering{arabic}


\section{Introduction}
\label{sec:Introduction}
 
The production of heavy quark-antiquark resonances (quarkonia) at hadron colliders
provides important information about the strong interaction, which is described by Quantum Chromodynamics (QCD).
Based on the assumption of factorization, the production process can be separated into two steps,
the generation of a heavy quark pair  $(Q\overline{Q})$ through the interaction of partons, 
and the subsequent hadronization of the quark pair into the quarkonium state.
The first step can be calculated within perturbative QCD, while the latter one is 
nonperturbative and needs to be described by phenomenological models. 
In the Colour Singlet Model~\cite{Carlson:1976cd, Donnachie:1976ue, Ellis:1976fj, Fritzsch:1977ay, Gluck:1978uia, Chang:1979nn, Baier:1981uk}, the intermediate $Q\overline{Q}$ state is assumed to be colourless, and has the same spin-parity quantum numbers as the final state quarkonium.
On the other hand, in the Colour Evaporation Model~\cite{Fritzsch:1977ay, Halzen:1977rs, Gluck:1977zm, Wu:1980mnt}, the probability of forming a specific quarkonium state is assumed to be independent of the
color of the $Q\overline{Q}$ pair.
In the nonrelativistic QCD (NRQCD) approach~\cite{Bodwin:1994jh, Cho:1995vh, Cho:1995ce}, intermediate $Q\overline{Q}$ states with all possible colour-spin-parity quantum numbers may evolve into the desired quarkonium. The transition probabilities, described by long-distance matrix elements (LDMEs) in NRQCD, are nonperturbative and are determined by fitting to experimental data. The LDMEs are assumed to be independent of the production mechanism and kinematics.

Although our understanding of hadronic production of heavy quarkonium has
improved significantly owing to four decades of
 theoretical and experimental efforts,
the underlying production mechanism is still not fully understood. For example, models
fail to  predict both cross-sections and polarisations of the measured $S$-wave heavy quarkonia states over the entire transverse momentum (\pt) and rapidity ($y$) range~\cite{Chen:2021tmf,Brambilla:2010cs,Chapon:2020heu}.
Furthermore, model calculations overestimate low-\pt ($\pt$ less than the $Q\overline{Q}$ invariant mass) cross-sections~\cite{Chen:2021tmf}, which dominate the \pt-integrated production.
The direct measurement of heavy quarkonium cross-sections at different colliding energies is a powerful tool to test production models and provides inputs to studies using heavy ion data~\cite{LHCb-PAPER-2014-015}.

At LHCb, the production of $\Upsilonres$ mesons has been studied via the $\mumu$ final state in $pp$ collisions at the centre-of-mass energies 
of $2.76, 7, 8$ and $13 \tev$~\cite{LHCb-PAPER-2013-066,LHCb-PAPER-2015-045,LHCb-PAPER-2018-002}.
In this analysis, a measurement of the production cross-sections of
$\OneS$, $\TwoS$ and $\ThreeS$ is performed, at the centre-of-mass energy of $\sqrt{s}=5 \tev$.
The data are collected by the LHCb detector in 2015 and correspond to an integrated luminosity of $9.13\pm0.18\invpb$~\cite{LHCb-PAPER-2014-047}.
The production cross-sections are measured
as functions of the transverse momentum and the rapidity of $\Upsilonres$ mesons, 
in the range $\pt<20 \gevc$ and $2.0<y<4.5$.
The result is used to update the nuclear modification factor $(R_{p\mathrm{Pb}})$ reported by LHCb for $\Upsilonres$ production in $p$Pb collisions at the nucleon-nucleon centre-of-mass energy of $\sqsnn=5\tev$~\cite{LHCb-PAPER-2014-015}.

\section{The \lhcb detector and event selection}
\label{sec:Detector}
The \lhcb detector~\cite{LHCb-DP-2008-001,LHCb-DP-2014-002} is a single-arm forward
spectrometer covering the \mbox{pseudorapidity} range $2<\eta <5$,
designed for the study of particles containing \bquark or \cquark
quarks. The detector includes a high-precision tracking system
consisting of a silicon-strip vertex detector surrounding the $pp$
interaction region~\cite{LHCb-DP-2014-001}, a large-area silicon-strip detector located
upstream of a dipole magnet with a bending power of about
$4{\mathrm{\,Tm}}$, and three stations of silicon-strip detectors and straw
drift tubes~\cite{LHCb-DP-2017-001}
placed downstream of the magnet.
The tracking system provides a measurement of the momentum, \ptot, of charged particles with
a relative uncertainty that varies from 0.5\% at low momentum to 1.0\% at 200\gevc.
Muons are identified by a
system composed of alternating layers of iron and multiwire
proportional chambers~\cite{LHCb-DP-2012-002}.

The online event selection is performed by a trigger~\cite{LHCb-DP-2012-004}, 
which consists of a hardware stage selecting events with at least one muon candidate with $\pt > 0.9 \gevc$ based on information from the 
muon systems, followed by a two-stage software stage, which applies a full event reconstruction.
In the software trigger, two muon candidates, each having $\pt>0.5\gevc$ and $p > 3 \gevc$, are required to form a good quality vertex with invariant mass $m(\mumu)>4.7\gevcc$. In between the two software stages, an alignment and calibration of the detector is performed in near real-time and their results are used in the second stage of the trigger~\cite{LHCb-PROC-2015-011}.
The same alignment and calibration information is propagated 
to the offline reconstruction, ensuring consistent and high-quality 
particle identification (PID) information between the trigger and 
offline software. The identical performance of the online and offline 
reconstruction offers the opportunity to perform physics analyses 
directly using candidates reconstructed in the trigger~\cite{LHCb-DP-2012-004,LHCb-DP-2016-001},
which the present analysis exploits. 

Candidate $\Upsilonres$ mesons are further selected offline, where two muon tracks are required to have  $\pt>0.65\gevc$, $p>10\gevc$ and $1.9<\eta<4.9$, and the invariant mass of the two muons is required to be in the range of $9.0<m(\mumu)<10.7\gevcc$.

Simulation is required to model the effects of the detector acceptance and the
imposed selection requirements.
In the simulation, $pp$ collisions are generated using
\pythia~\cite{Sjostrand:2007gs,*Sjostrand:2006za} 
with a specific \lhcb configuration~\cite{LHCb-PROC-2010-056}.
Decays of unstable particles
are described by \evtgen~\cite{Lange:2001uf}, in which final-state radiation is generated using \photos~\cite{davidson2015photos}.
The interaction of the generated particles with the detector, and its response,
are implemented using the \geant
 toolkit~\cite{Allison:2006ve, *Agostinelli:2002hh} as described in
 Ref.~\cite{LHCb-PROC-2011-006}.

\section{Cross-section determination}
The double-differential cross-section times the $\Upsilonres\to\mumu$ branching fraction ($\mathcal{B}$) of $\Upsilonres$ production
in a given (\pt,$y$) bin is defined as
\begin{equation}
\label{eq:cs}
	\mathcal{B}(\emph{\Upsilonres}\to\mumu)\times\frac{\deriv^2\sigma}{\deriv \emph{p}_T \deriv \emph{y}}=
	\frac{\emph{N}(\emph{\Upsilonres}\to\mumu)}{\lum\times\etot\times\Delta\emph{p}_T\times\Delta \emph{y}},
\end{equation}
where $N(\Upsilonres\to\mumu)$ is the yield of $\Upsilonres\to\mumu$ signal decays,
\etot is the total efficiency in a (\pt,$y$) bin,
\lum is the integrated luminosity,
and $\Delta\pt$ and $\Delta y$ are the interval widths of the $\Upsilonres$ transverse momentum and rapidity, respectively.

The $\Upsilonres$ yield in each (\pt,$y$) bin
is determined with an extended unbinned maximum-likelihood fit
to the invariant-mass distribution of the  $\Upsilonres$ candidates, denoted as $m_\Upsilonres$ in the following. Four components are used to describe the distributions, the three $\Upsilonres$ signals and a combinatorial background. 
The mass distribution for each of the $\OneS$, $\TwoS$ and $\ThreeS$ states
is described by a Crystal Ball (CB) function~\cite{Skwarnicki:1986xj}. The differences between the means and the ratios of the widths of these CB functions are fixed, respectively, to the differences and the ratios of the world-average  $\Upsilonres(nS)$ masses~\cite{PDG2022:2022ynf}.
The CB function parameters $\alpha$ and \emph{n}, which describe the radiative tail of the $\Upsilonres$ mass distribution, are fixed to the values of $\alpha=2$ and $\emph{n}=1$, following the same approach of the previous analyses~\cite{LHCb-PAPER-2015-045,LHCb-PAPER-2018-002}.
The mass distribution of the combinatorial background contribution is described by an exponential function.

\begin{figure}[!htb]
	\centering
	\includegraphics[width=0.8\linewidth]{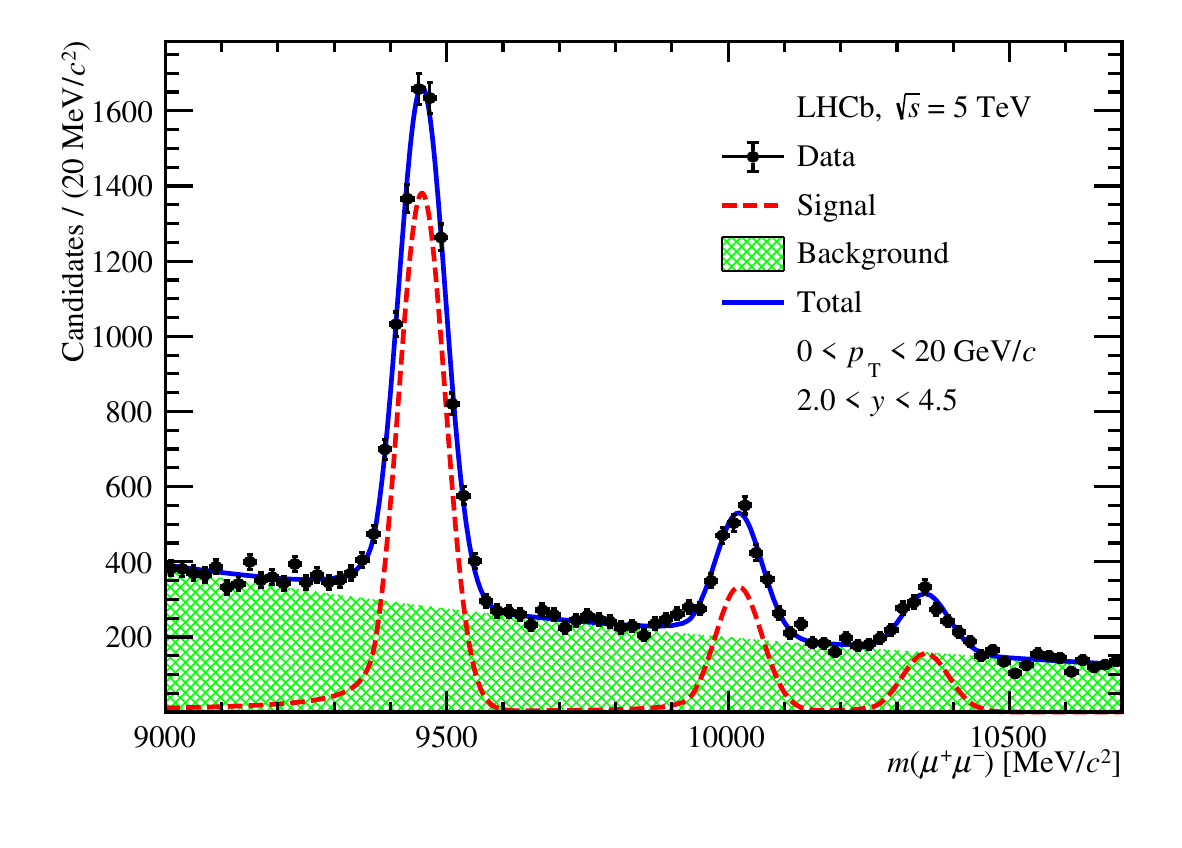}
	\caption{Invariant mass distribution of $\Upsilonres$ candidates in the kinematic range $\pt\in[0,20]\gevc$ and $y\in[2.0,4.5]$. The result of the fit with three Crystal Ball functions for the signal plus an exponential function for the background is also shown. 
	}
	\label{fig:fit_all}
\end{figure}

The invariant mass distribution of all candidates for 
$\pt\in[0,20]\gevc$ and $y\in[2.0,4.5]$ is shown in Fig.~\ref{fig:fit_all}, with the fit result superimposed.
The yields of $\OneS$, $\TwoS$ and $\ThreeS$ signals are $7560\pm120$, $1900\pm70$ and $930\pm60$, respectively.

The total efficiency \etot is expressed as the product of four components: the geometrical acceptance, $\epsilon_\text{Acc}$, the reconstruction and selection efficiency, $\epsilon_\text{Rec\&Sel}$, the muon identification efficiency, $\epsilon_\text{PID}$, and the trigger efficiency, $\epsilon_\text{Trg}$,  as
\begin{equation}
	\etot=\epsilon_\text{Acc}\times\epsilon_\text{Rec\&Sel}\times\epsilon_\text{Pid}\times\epsilon_\text{Trg}.
\end{equation}

These efficiencies in general depend on the $\Upsilonres$ polarisation.
The measurement of the $\Upsilonres$ polarisation by \lhcb at $\sqs=7$ and $8 \tev$~\cite{LHCb-PAPER-2017-028} in $pp$ collisions indicates that the three polarisation parameters $\lambda_{\theta}$, $\lambda_{\theta\phi}$ and $\lambda_{\phi}$ are consistent with zero. 
Therefore, the efficiencies are determined in each (\pt, $y$) bin for the $\OneS$, $\TwoS$ and $\ThreeS$ mesons, assuming the $\Upsilonres$ states are unpolarised. 
Variations of cross-sections due to polarisation are studied for different scenarios of the polarisation parameter $\lambda_{\theta}$, as the dependence of the efficiency on the $\lambda_{\theta\phi}$ and $\lambda_{\phi}$ parameters is smaller~\cite{LHCb-PAPER-2011-003}.

The quantity $\epsilon_\text{Acc}$ is calculated using the simulation sample without a prior geometrical acceptance requirement on the muons.
The efficiency $\epsilon_\text{Rec\&Sel}$ is determined  using full simulation samples for which the distributions  of charged track multiplicity of $pp$ collisions and  $\Upsilonres$ rapidity are  corrected to  match data.
The efficiency $\epsilon_\text{PID}$ is calculated using the single muon PID efficiency in bins of $(p,\eta)$ weighted by the muon $(p,\eta)$ distribution in simulation samples. The single-muon PID efficiency is obtained using $\jpsi\to\mumu$ decays in calibration data~\cite{LHCb-DP-2013-001,LHCb-DP-2018-001}. 
The  efficiency $\epsilon_\text{Trg}$ is determined on simulated events.

\section{Systematic uncertainties}
Several sources of systematic uncertainty are expected to affect the cross-section measurements:
determination of the signal yields, efficiency calculations and computation of the integrated luminosity.
The efficiency-related uncertainties are determined with independent simulation samples for each $\Upsilonres(nS)$ meson, and are expected to be similar for the three states.
They are reported in Table~\ref{table:summary} and discussed in detail in the following.  The total uncertainty is calculated as the quadrature sum of these uncertainties.

\begin{table}[!htb]
	\centering
	\caption{Relative systematic uncertainty ($\%$) from various sources. The ranges correspond to results measured in each $(\pt,y)$ bin of $\Upsilonres$ mesons. For some systematic uncertainties, the results are correlated among different $(\pt,y)$ bins as noted in the last column.}
	\label{table:summary}
	\begin{tabular}{l|lllll}
		\hline
		        & Source       &$\OneS$       &$\TwoS$       &$\ThreeS$     &Comment\\
		\hline
		Signal yields                           & Fit model    &2.5 &2.8 &3.1 &Correlated\\
		\hline
		\multirow{7}{*}{Efficiency $\epsilon$}  & Trigger   &0.7--1.9 &0.7--2.1 &0.7--1.9 &Correlated\\
	                                            & Track reconstruction  &1.7--3.1 &1.7--3.2 &1.7--3.4 &Correlated\\
	                                            & MuonID    &0.5--5.3 &0.5--5.1 &0.5--5.2 &Correlated\\
	                                            & Kinematic spectrum    &0.0--4.9 &0.0--4.5 &0.0--6.3 &Uncorrelated\\
	                                            & Final-state radiation    &1.0 &1.0 &1.0 &Correlated\\
	                                            &Simulation sample size    &0.5--4.5 &0.4--3.3 &0.4--3.4 &Uncorrelated\\
\hline
		 Luminosity                   &      & 2.0&2.0 &2.0 &Correlated\\
		 \hline
	\end{tabular}
\end{table}

The systematic uncertainty from the determination of the signal yields is affected by the invariant-mass fit model and is  estimated using pseudoexperiments. The pseudoexperiments employ a mixture of $\OneS$ signal in the full simulation sample and
background generated according to the background model fitted in data. In the pseudoexperiments, the  relative fraction of the two components is chosen such that the level of background matches the data in the $\Upsilonres$ signal region.
This signal-plus-background sample is fitted with the same model as for the data.
The systematic uncertainty is estimated as the relative difference between the signal yield obtained from the fit and that of the input. This study is repeated for the $\OneS$, $\TwoS$ and  $\ThreeS$ states, with different fractions of background, yielding $2.5\%$, $2.8\%$ and $3.1\%$ as the systematic uncertainty for the three states, respectively. The small difference between the uncertainties for the three states is attributed to different background levels under each signal peak.

Systematic uncertainties on both hardware and software trigger efficiency are studied in this analysis.
To validate systematic effects on the hardware trigger, the trigger efficiency is determined for $\mup$ and $\mun$ individually and is then used to calculate the efficiency for $\Upsilonres$ signals following the method described in Ref.~\cite{LHCb-PAPER-2021-020}.
The difference between the trigger efficiencies obtained on data and simulation is quoted as the systematic uncertainty on the hardware-trigger efficiency.
The systematic uncertainty, estimated to be in the range 0.0--1.9\%, depends on the $(\pt,y)$ bin of the $\Upsilonres$ mesons and is assumed to be correlated among the bins.
For the software trigger, alternative efficiency values are determined using a subset of events triggered independently of the $\Upsilonres$ signals~\cite{LHCb-DP-2019-001}.
Within this sample, the fraction of $\Upsilonres$ signal events that passes the software requirement is quoted as the trigger efficiency. The efficiency is calculated for both data and simulation samples in $(\pt,y)$ bins of the  $\Upsilonres$ mesons, and their difference is taken as the systematic uncertainty. 

The track reconstruction efficiency is evaluated for each muon using calibrated simulation and data samples~\cite{LHCb-DP-2013-002}, and their difference is used to correct the $\epsilon_\text{Rec\&Sel}$ efficiency of simulated $\Upsilonres$ decays.
A systematic uncertainty of $0.8\%$ per track is assigned to account for the dependency of the correction on the event multiplicity variable that is used to weight simulation to match data.
Furthermore, an uncertainty is associated with the correction due to the limited size of the calibration samples, which is propagated to the $\Upsilonres$ cross-section measurement using a large number of pseudoexperiments.

The muon identification efficiency is affected by
the statistical fluctuation of the PID calibration constants originating from the limited size of calibration samples 
and from the choice of the $(p,\eta)$ binning scheme for muons.
The statistical uncertainties in the PID efficiencies are propagated to the $\Upsilonres$ cross-sections using pseudoexperiments. The variation is found to be below $1\%$ and is therefore neglected.
The systematic uncertainty due to the kinematic binning scheme of the calibration sample
is studied by measuring the PID efficiency using alternative binning schemes.
The difference between the alternative efficiency and the default one is quoted as the systematic uncertainty.

The kinematic distributions of the $\Upsilonres$ mesons in data and simulation samples could be different, which may cause a mis-modelling in the underlying detection efficiency.
Efficiencies measured in $(\pt,y)$ intervals help to reduce the discrepancy, however the $(\pt,y)$ bin size can cause systematic effects in the efficiency determination.
To check for possible residual effects,
the $y$ distributions of the $\Upsilonres$ states in simulation  are reweighted
to match those in data,
for which the background has been subtracted using the $\sPlot$ technique~\cite{Pivk:2004ty}.
In this analysis, the $y$ distribution is studied as the  efficiencies are found to have only a weak dependence on the $\pt$ of the $\Upsilonres$ candidates. 
All the efficiencies are re-calculated with the reweighted $y$ spectrum in the simulation sample,
and relative differences between the new results and the default are taken as
systematic uncertainties.

Due to the missing energy caused by final-state radiation that is not reconstructed,
some signal events are removed by the requirement on the $\mumu$ invariant mass.
The efficiency of this requirement is studied using simulation samples.
A systematic uncertainty of $1.0\%$ is assigned to take into account possible differences between data and simulated $\Upsilonres$ decays, following the method of Ref.~\cite{LHCb-PAPER-2018-002}.

The limited size of the simulation samples used to determine the efficiencies is a source of systematic uncertainty. The relative statistical uncertainties in the efficiencies are propagated to the cross-section measurements.
The integrated luminosity has a relative uncertainty of $2.0\%$~\cite{LHCb-PAPER-2014-047},
which is propagated to the final results as a systematic uncertainty.

\section{Results}
\subsection{Cross-sections}
The double-differential cross-sections times the branching fraction of 
$\Upsilonres\to\mumu$ are shown in Fig. \ref{fig:res_1S} and the numeric results are listed in Tables~\ref{table1S}, \ref{table2S} and \ref{table3S} in Appendix~\ref{sec:table} for $\OneS$, $\TwoS$ and $\ThreeS$, respectively.
The cross-section times branching fraction as a function of $\pt$ and integrated over $y$ or vice-versa are shown in Fig.~\ref{fig:res_int}, and listed in Tables~\ref{table_pt} and~\ref{table_y} in Appendix~\ref{sec:table} for the three $\Upsilonres$ states.
The  change of $\Upsilonres$ cross-sections for alternative polarisation relative to the zero polarisation is studied for $\lambda_{\theta} = -1, 0.1$ and $+1$ in the helicity frame, which is reported in Appendix~\ref{appendixB}. The result for $\lambda_{\theta} = \pm1$ corresponds to the extreme transverse and longitudinal polarisation and $\lambda_{\theta}=0.1$ is the approximation of \lhcb measurement~\cite{LHCb-PAPER-2017-028}.
Results for other $\lambda_{\theta}$ values can be interpolated.

\begin{figure}[!htb]
	\centering
	\includegraphics[width=0.45\linewidth]{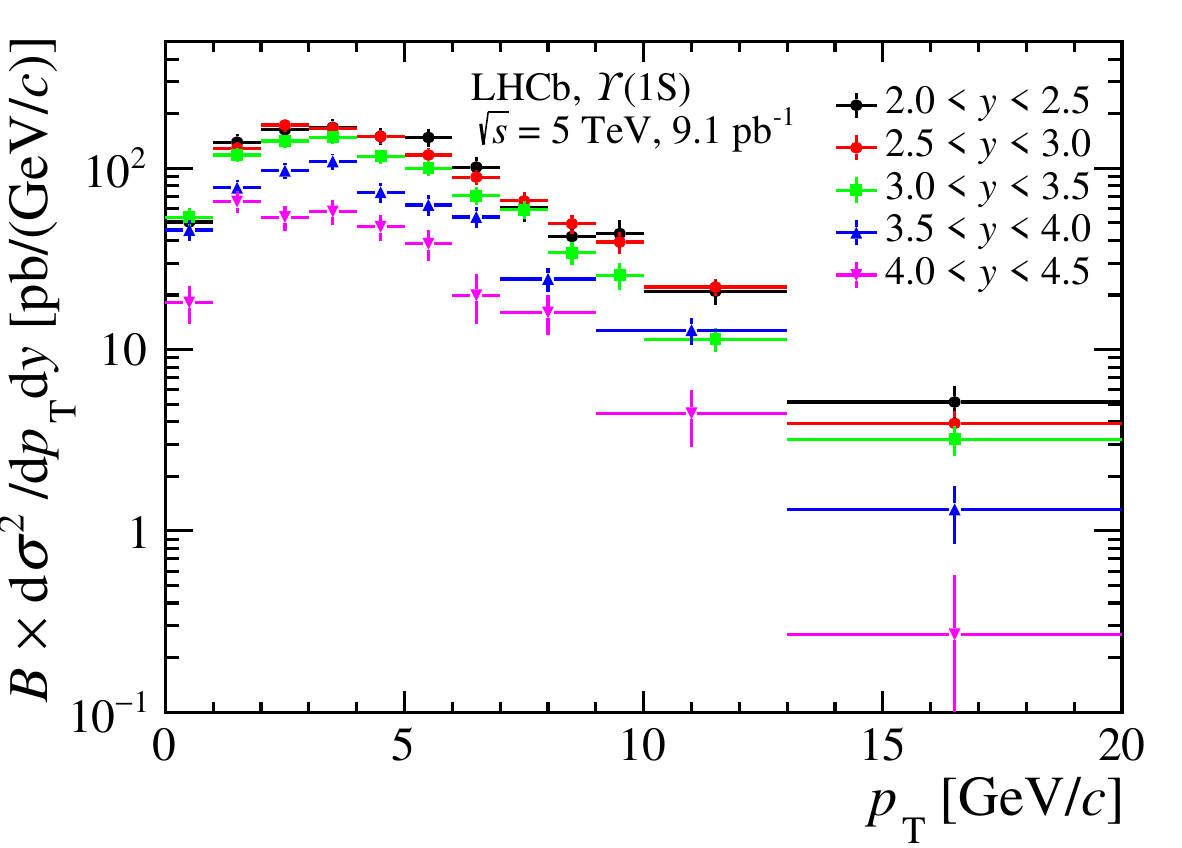}
	\includegraphics[width=0.45\linewidth]{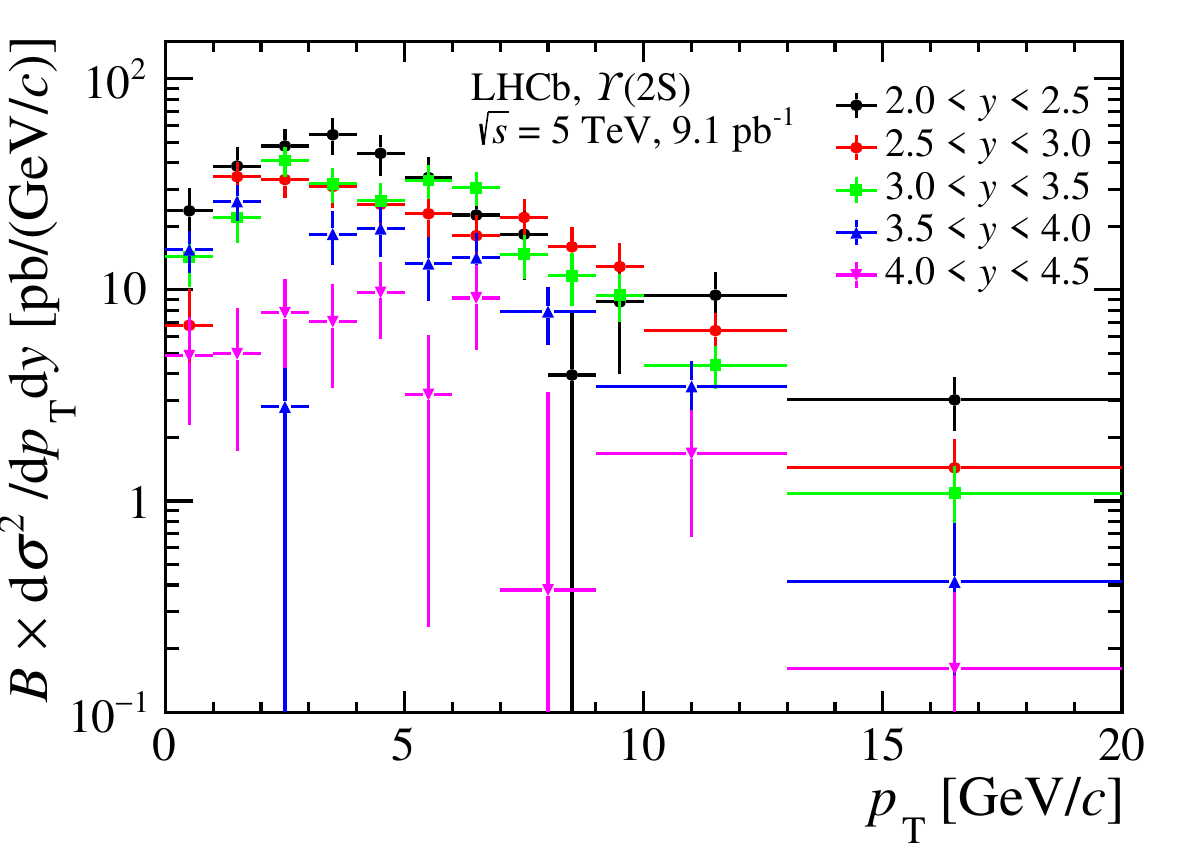}
	\includegraphics[width=0.45\linewidth]{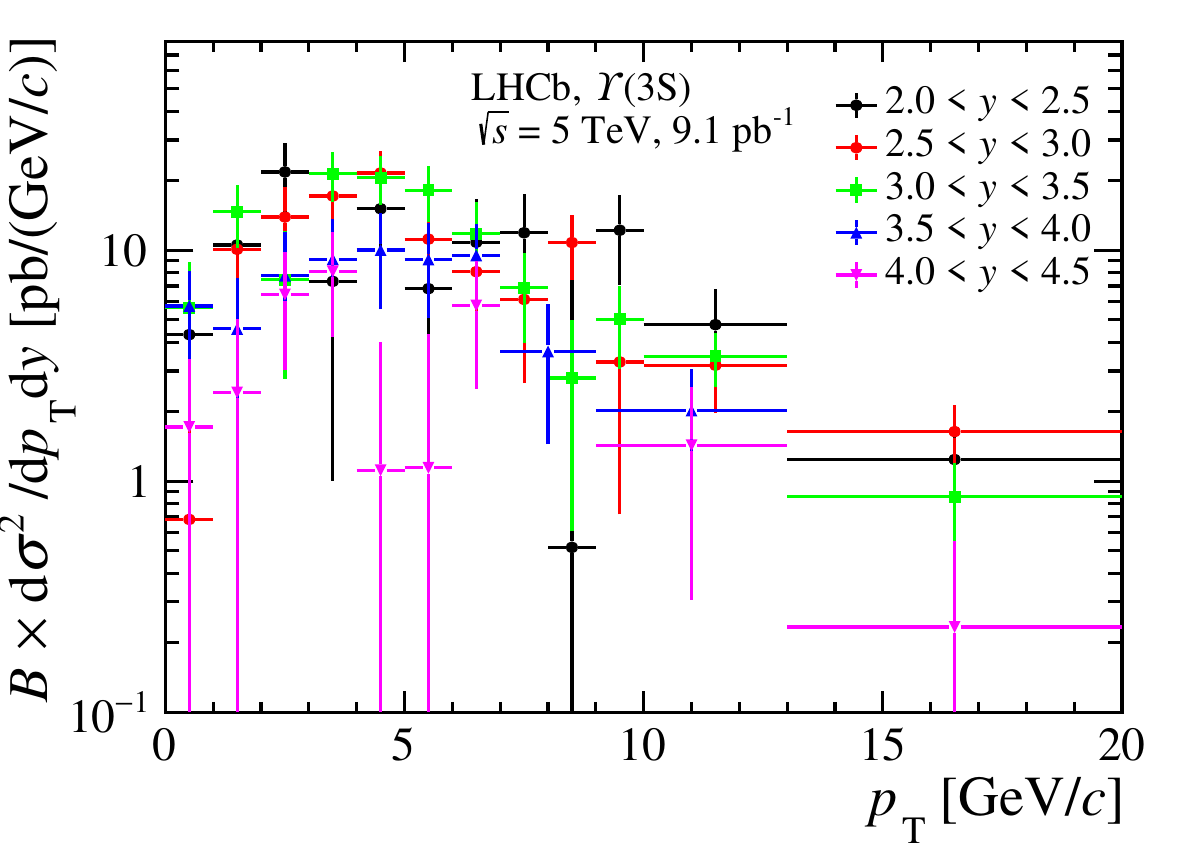}
	\caption{Double-differential cross-section times $\Upsilonres\to\mumu$ branching fraction as a function of $\pt$ in bins of $y$ for (upper left) $\OneS$, (upper right) $\TwoS$ and (bottom) $\ThreeS$ mesons. Statistical and systematic uncertainties are added in quadrature.}
	\label{fig:res_1S}
\end{figure}

\begin{figure}[!htb]
    \centering
    \includegraphics[width=0.45\linewidth]{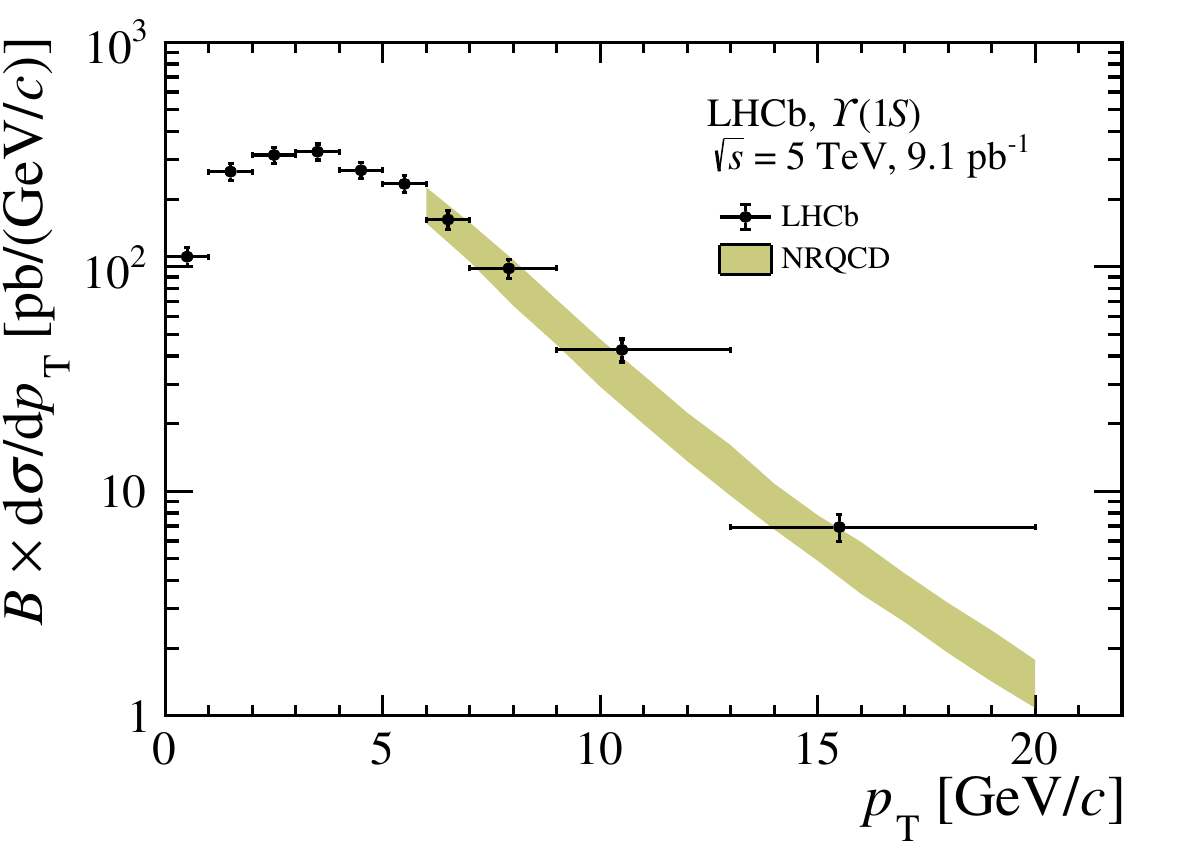}
    \caption{Differential cross-section of $\OneS$ mesons as a function of $\pt$, compared with the NRQCD prediction of Ref.~\cite{Feng:2015wka}. }
    \label{fig:compare to theory}
\end{figure}

The differential cross-section times branching fraction as a function of $\pt$ for $\OneS$ mesons for $y$ integrated over $2.0$--$4.5$  are compared with the NRQCD predictions~\cite{Feng:2015wka} in Fig.~\ref{fig:compare to theory}. The data points are plotted at the mean $\pt$ of each bin. Good agreement is observed between NRQCD predictions and the measurement of this analysis for $\pt > 5 \gevc$. Another NRQCD calculation that tends to describe high-\pt data at rapidity $y\sim0$ is found to overestimate the LHCb result~\cite{Han:2014kxa}. The difference between the two NRQCD calculations is due to different values of LDMEs used in their analyses.

\begin{figure}[!htb]
    \centering
	\hfil
    \begin{minipage}[t]{0.45\linewidth}
        \centering
        \includegraphics[width=\linewidth]{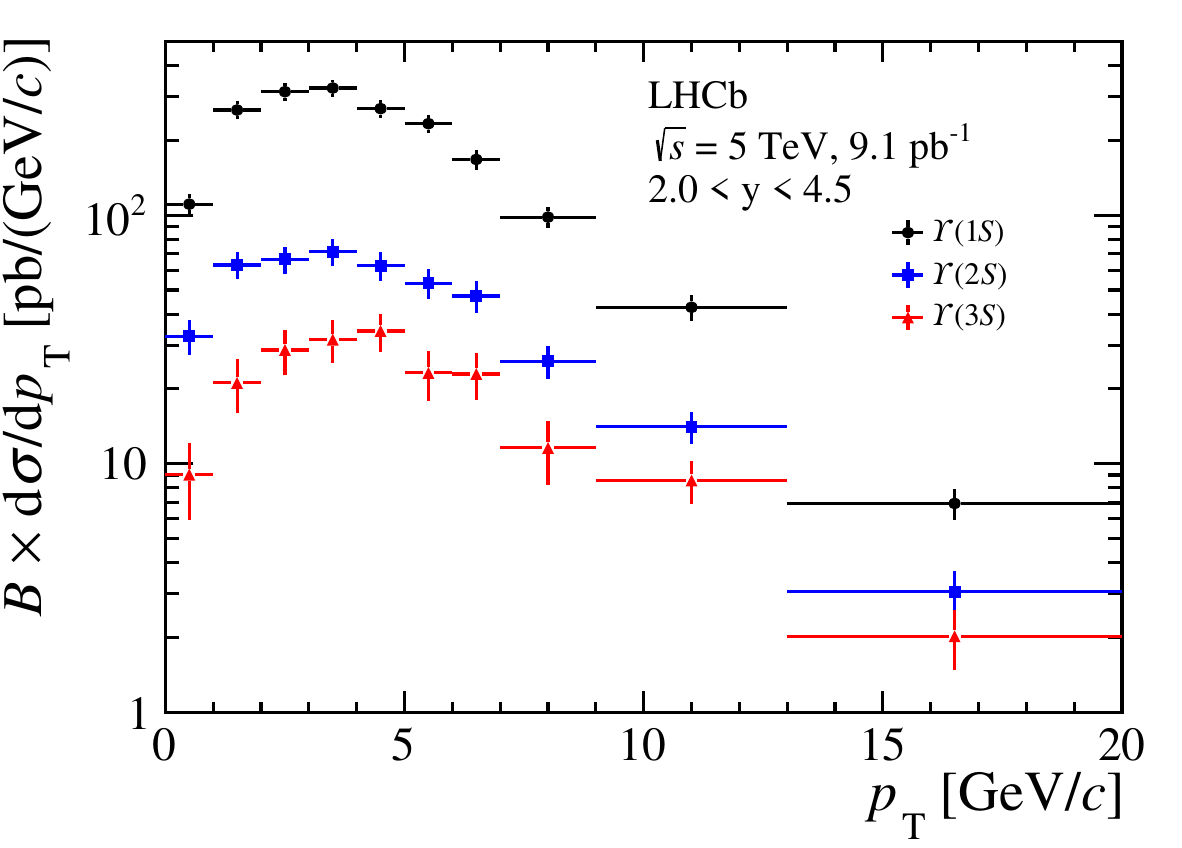}
    \end{minipage}
    \begin{minipage}[t]{0.45\linewidth}
        \centering
        \includegraphics[width=\linewidth]{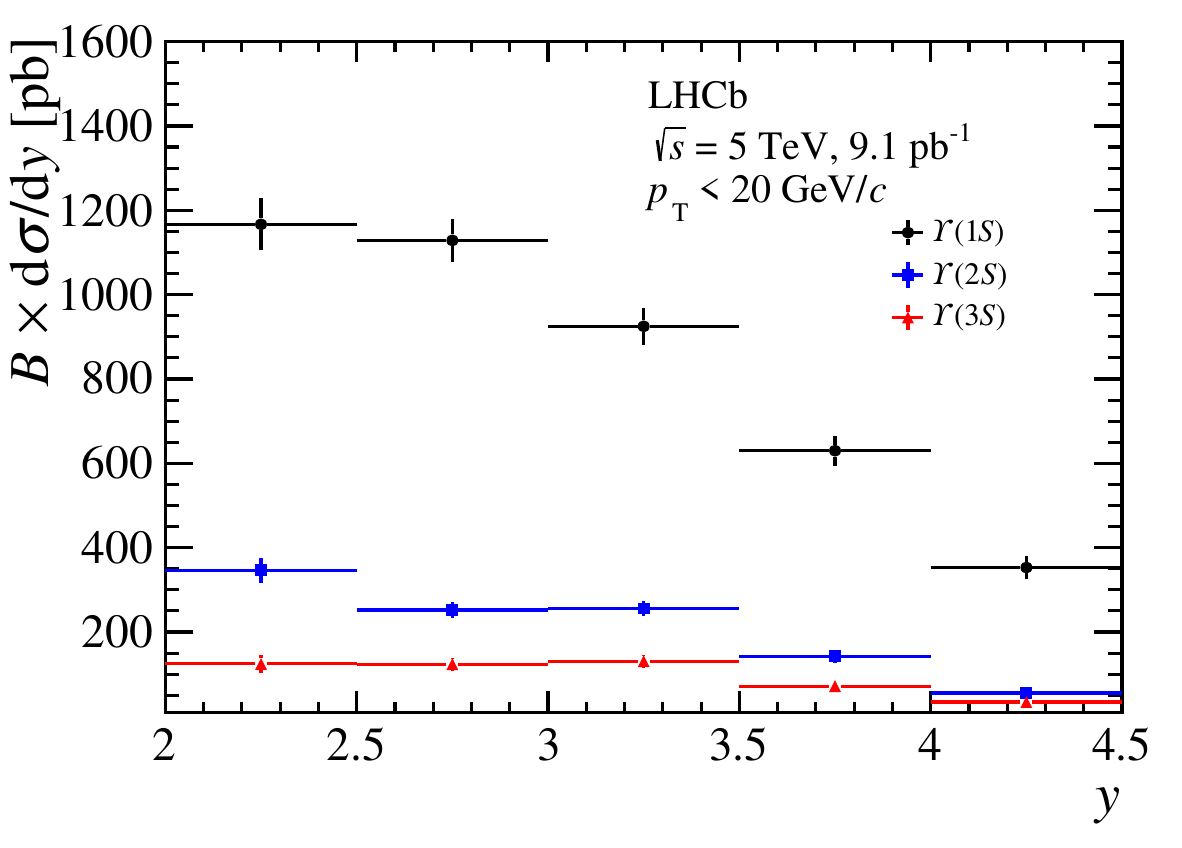}
    \end{minipage}
    \hfil
	\caption{Differential cross-section times $\Upsilonres\to\mumu$ branching fraction  (left) as a function of $\pt$ for $y$ integrated from 2.0 to 4.5 and (right) as a function of $y$ for $\pt$ integrated from 0 to 20\gevc, for the $\OneS$, $\TwoS$ and $\ThreeS$ states. Statistical and systematic uncertainties are added in quadrature.}
    \label{fig:res_int}
\end{figure}

\begin{figure}[!htb]
	\centering
	\includegraphics[width=0.5\linewidth]{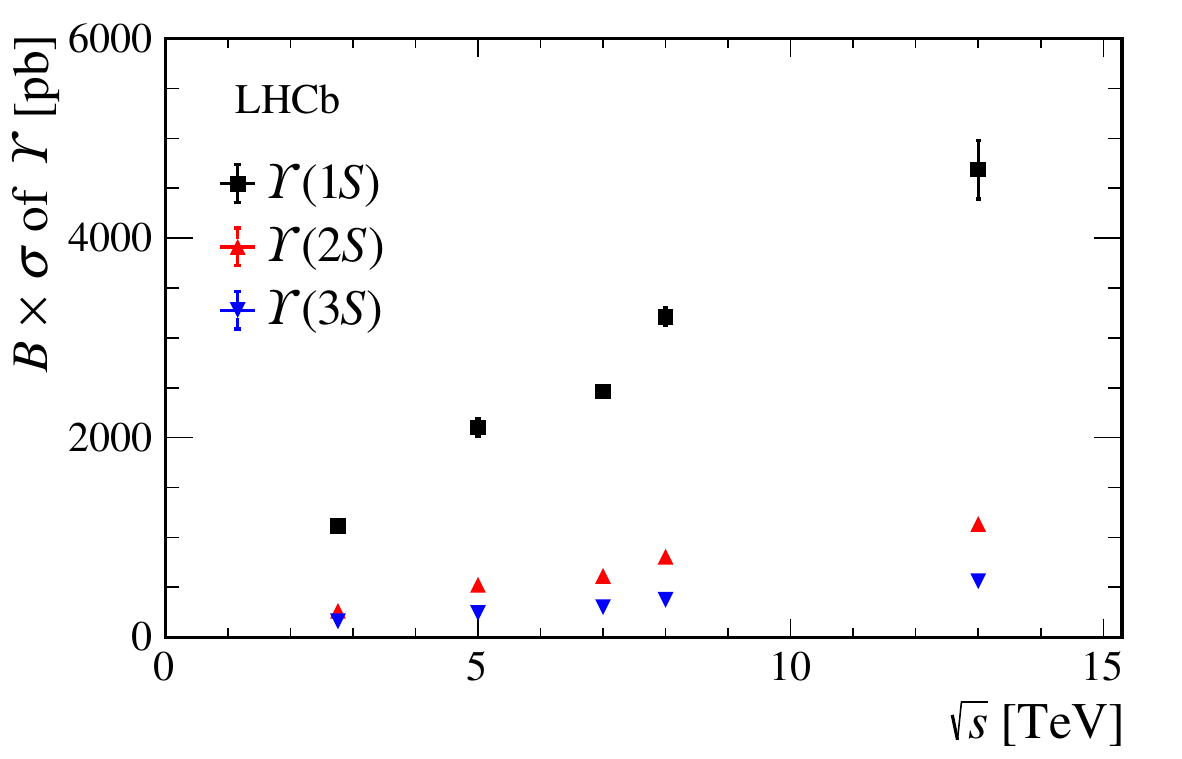}
	\caption{Production cross-sections in the range $\pt < 15 \gevc$ and $2.0 < y < 4.5$ multiplied by $\Upsilonres\to\mumu$ branching fractions as a function of the centre-of-mass energy of $pp$ collisions for (black boxes) $\OneS$,  (red upward triangles) $\TwoS$ and  (blue downward triangles) $\ThreeS$ states. Statistical and systematic uncertainties are added in quadrature.}
	\label{fig:Xsec_energy}
\end{figure}

The double-differential cross-sections integrated over the kinematic range of \mbox{$\pt<20 \gevc$} and $2.0<y<4.5$ are
\begin{eqnarray}
  \sigma(\OneS) \times\mathcal{B}(\emph{\OneS}\to\mumu)&=&2101\pm33\stat\pm83\syst\pb,\nonumber \\ 
  \sigma(\TwoS) \times\mathcal{B}(\emph{\TwoS}\to\mumu)&=&\phantom0 526\pm20\stat\pm21\syst\pb,\nonumber\\ 
  \sigma(\ThreeS) \times\mathcal{B}(\emph{\ThreeS}\to\mumu)&=&\phantom0 242\pm16\stat\pm10\syst\pb.\nonumber
\end{eqnarray}
The total cross-section in the LHCb acceptance as a function of $pp$ centre-of-mass energy is shown in Fig.~\ref{fig:Xsec_energy} for $\OneS$, $\TwoS$ and $\ThreeS$ mesons. All three distributions show approximately a linear rise as a function of $\sqs$ in the range $2.76$--$13\tev$.

\subsection{Cross-section ratios}
\begin{figure}[!htb]
    \centering
    \hfil
    \begin{minipage}[t]{0.45\linewidth}
        \centering
        \includegraphics[width=\linewidth]{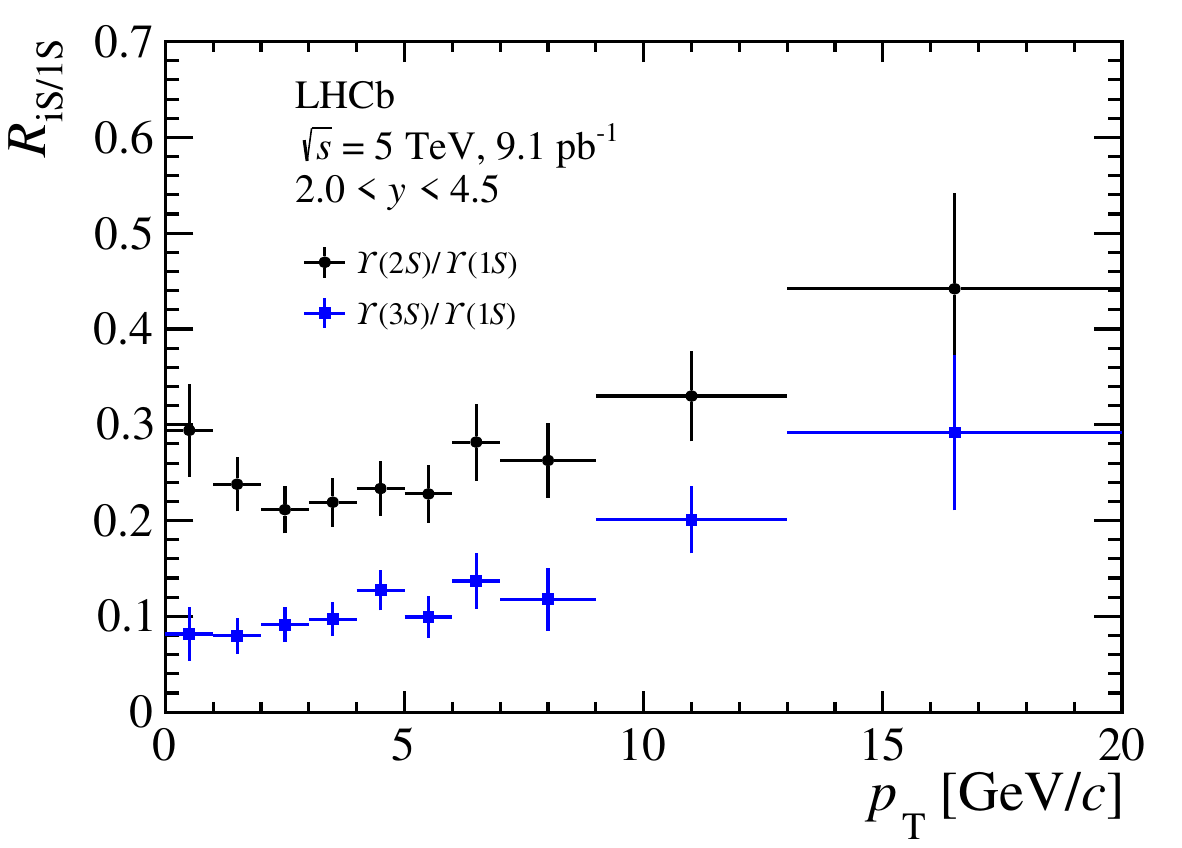}
    \end{minipage}
    \begin{minipage}[t]{0.45\linewidth}
        \centering
        \includegraphics[width=\linewidth]{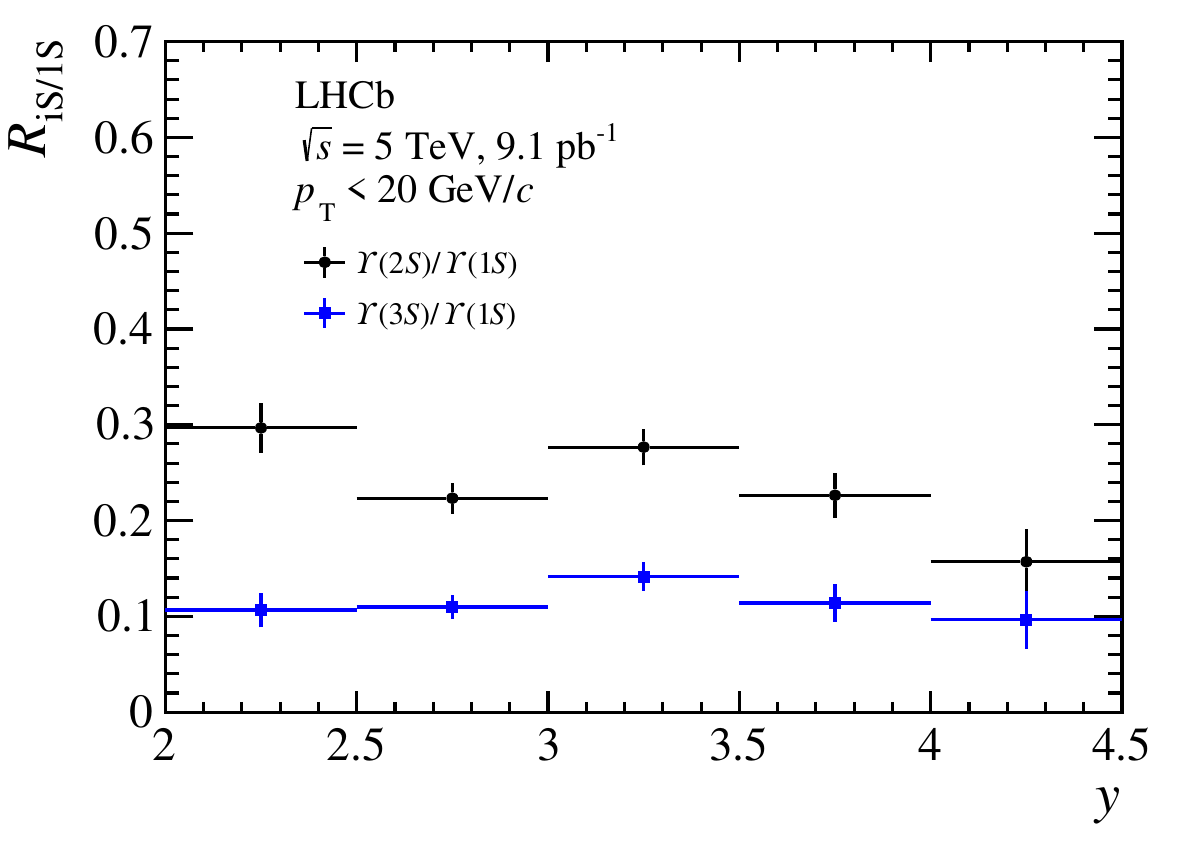}
    \end{minipage}
    \caption{Ratios of the $\TwoS$ or $\ThreeS$ production cross-section over that of the $\OneS$ state, (left) as a function of  $\pt$ for $2.0<y<4.5$ and (right)  as a function of $y$ for $\pt<20\gevc$. Statistical and systematic uncertainties are added in quadrature.}
    \label{fig:ratio_state}
\end{figure}

The cross-section ratios between the different $\Upsilonres$ states are calculated as:
\begin{eqnarray}
    R_{2S/1S} &\equiv& \frac{\frac{d\sigma_{\TwoS}}{dy(\pt)}\times\mathcal{B}(\emph{\TwoS}\to\mumu)}{\frac{d\sigma_{\OneS}}{dy(\pt)}\times\mathcal{B}(\emph{\OneS}\to\mumu)},\\
    R_{3S/1S} &\equiv& \frac{\frac{d\sigma_{\ThreeS}}{dy(\pt)}\times\mathcal{B}(\emph{\ThreeS}\to\mumu)}{\frac{d\sigma_{\OneS}}{dy(\pt)}\times\mathcal{B}(\emph{\OneS}\to\mumu)},
\end{eqnarray}
and the results are shown in Fig.~\ref{fig:ratio_state}.
The ratios increase slightly with \pt but 
show no obvious dependence on $y$.

In the evaluation of the cross-section ratios,
the statistical and systematic uncertainties due to the limited simulation sample sizes are assumed to be uncorrelated between the two $\Upsilonres$ states and, hence, cancel in the ratios.
The systematic uncertainties due to the invariant mass  fit model, and those originating from the trigger as well as tracking efficiencies are assumed to be fully correlated between cross-sections of two $\Upsilonres$ states.

Using the $\Upsilonres$  cross-sections measured by LHCb at $\sqs=13 \tev$~\cite{LHCb-PAPER-2018-002}, ratios of cross-sections  at $\sqs=13 \tev$ and $\sqs=5 \tev$ are determined. The results as a function of $\pt$ and those as a function of $y$ are shown in Fig.~\ref{fig:ratio_energy} and are summarized in Tables~\ref{tablenE_pt} and \ref{tablenE_y} in Appendix~\ref{sec:table} , from which it is seen that the ratio increases from about 2 at low \pt to about 4 at high \pt, but is almost independent of $y$. 

\begin{figure}[!htb]
    \centering
    \hfil
    \begin{minipage}[t]{0.45\linewidth}
        \centering
        \includegraphics[width=\linewidth]{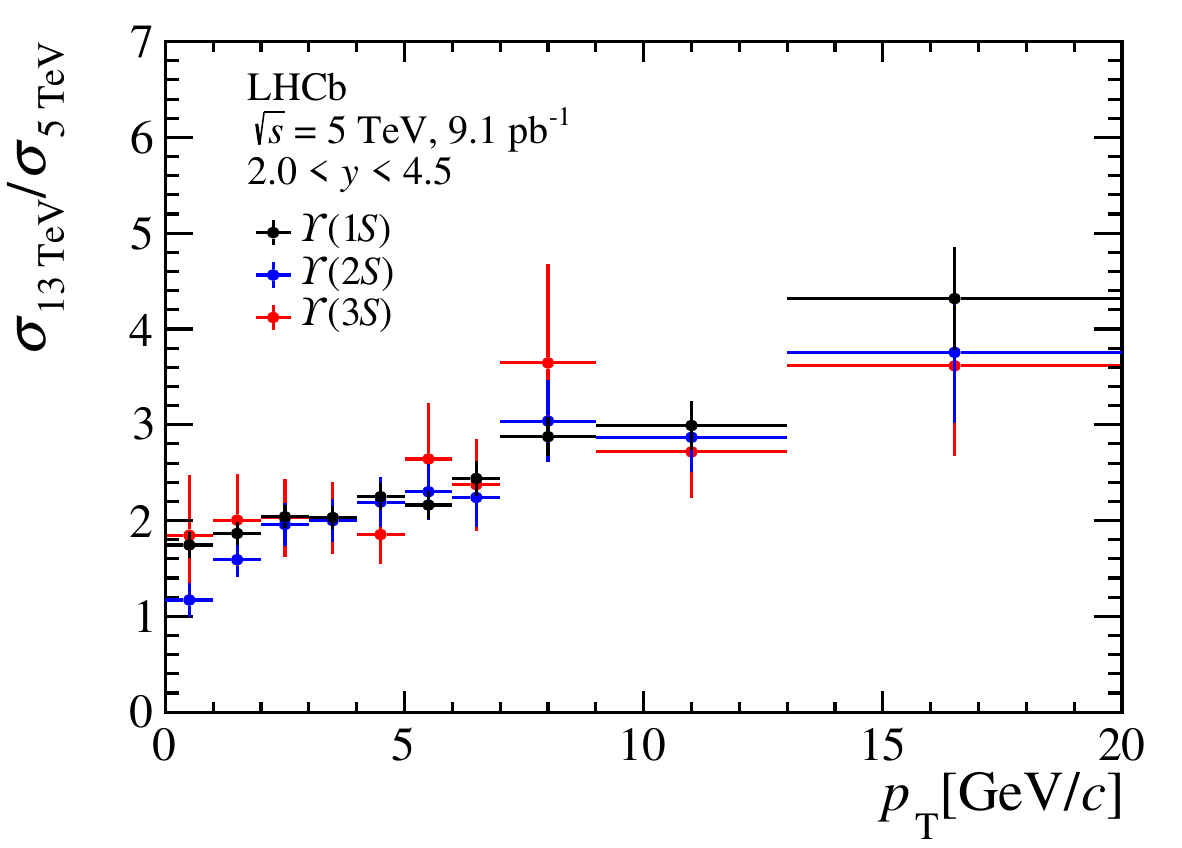}
    \end{minipage}
    \begin{minipage}[t]{0.45\linewidth}
        \centering
        \includegraphics[width=\linewidth]{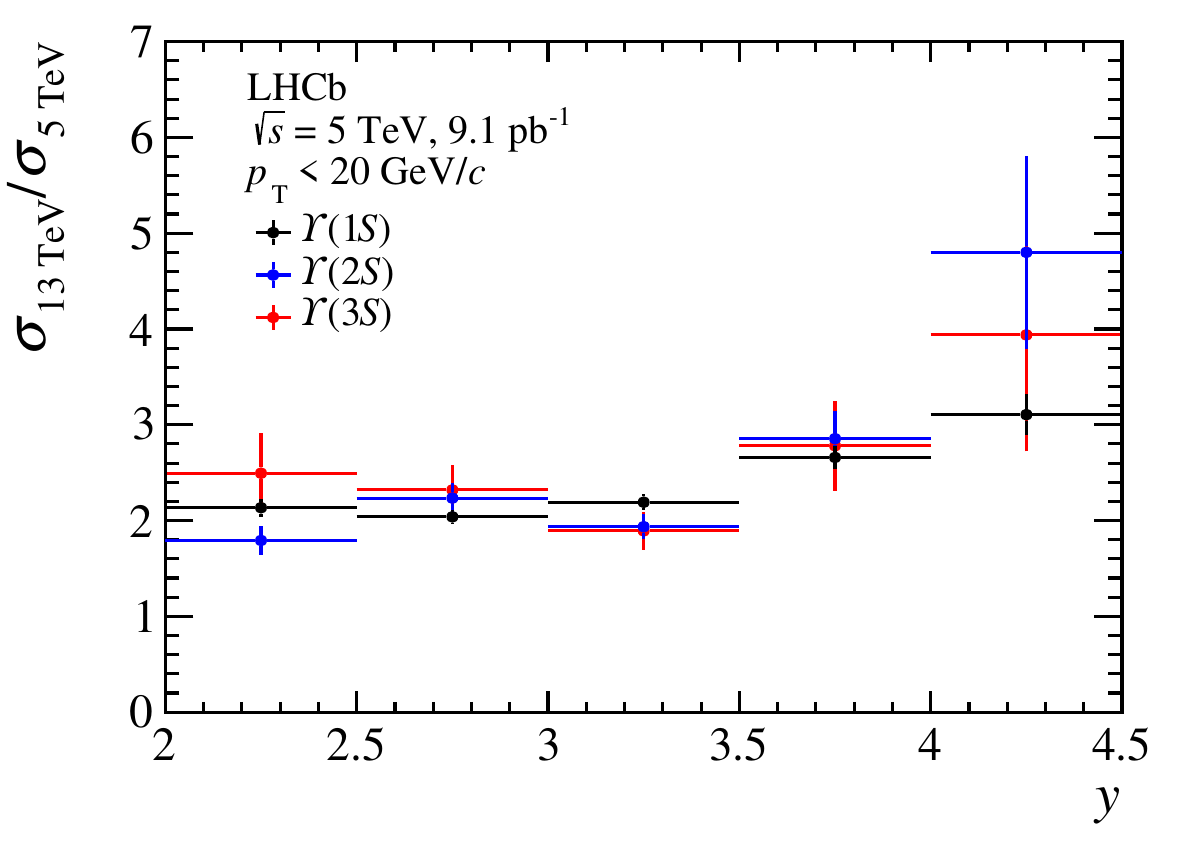}
    \end{minipage}
    \caption{Ratios of $\Upsilonres$ production cross-sections  measured at $\sqs=13\tev$ over those measured at $ \sqs=5\tev$ (left) as a function of $\pt$ for $2.0<y<4.5$ and (right) as a function of $y$ for $\pt<20\gevc$.
    The statistical and systematic uncertainties are added in quadrature.
    }
    \label{fig:ratio_energy}
\end{figure}

For the cross-section ratios between measurements at different $\sqs$, the systematic uncertainties due to the invariant mass fit model and final-state radiation are considered to be fully correlated. The systematic uncertainties originating from uncertainties of trigger, muon identification, tracking efficiencies and luminosity are assumed to be partially correlated.

\subsection{Update of the {\boldmath $\OneS$} nuclear modification factor}
To quantify nuclear effects for particle production in heavy-ion collisions, the nuclear modification factor
\begin{equation}
    R_{p{\rm Pb}} \equiv \frac{\sigma_{p\rm{Pb}}(\sqsnn)}{A \times \sigma_{pp}(\sqsnn)}
\end{equation}
is determined, where $A=208$ is the mass number for Pb, and $\sigma_{p\rm{Pb}}$ and $\sigma_{pp}$ are the cross-sections for proton-lead ($p$Pb) and $pp$ collisions at the same centre-of-mass energy per nucleon pair.
Using a previous measurement of $ R_{p{\rm Pb}}$ for $\OneS$ in $p$Pb collisions at $\sqsnn=5\tev$~\cite{LHCb-PAPER-2014-015}, the $\OneS$ production cross-section in $pp$ collisions at $\sqs=5\tev$
is obtained by an interpolation of LHCb measurements at $\sqs=2.76, 7$ and $8\tev$~\cite{LHCb-PAPER-2013-016,LHCb-PAPER-2011-036, LHCb-PAPER-2013-066}, as
\begin{equation*}
    \sigma(\OneS)\times\mathcal{B}(\emph{\OneS}\to\mumu)=1.12\pm0.11\nb,
\end{equation*}
in the range $\pt<15\gevc, 2.5<y<4.0$.

\begin{figure}[!htb]
    \centering
    \includegraphics[width=0.6\linewidth]{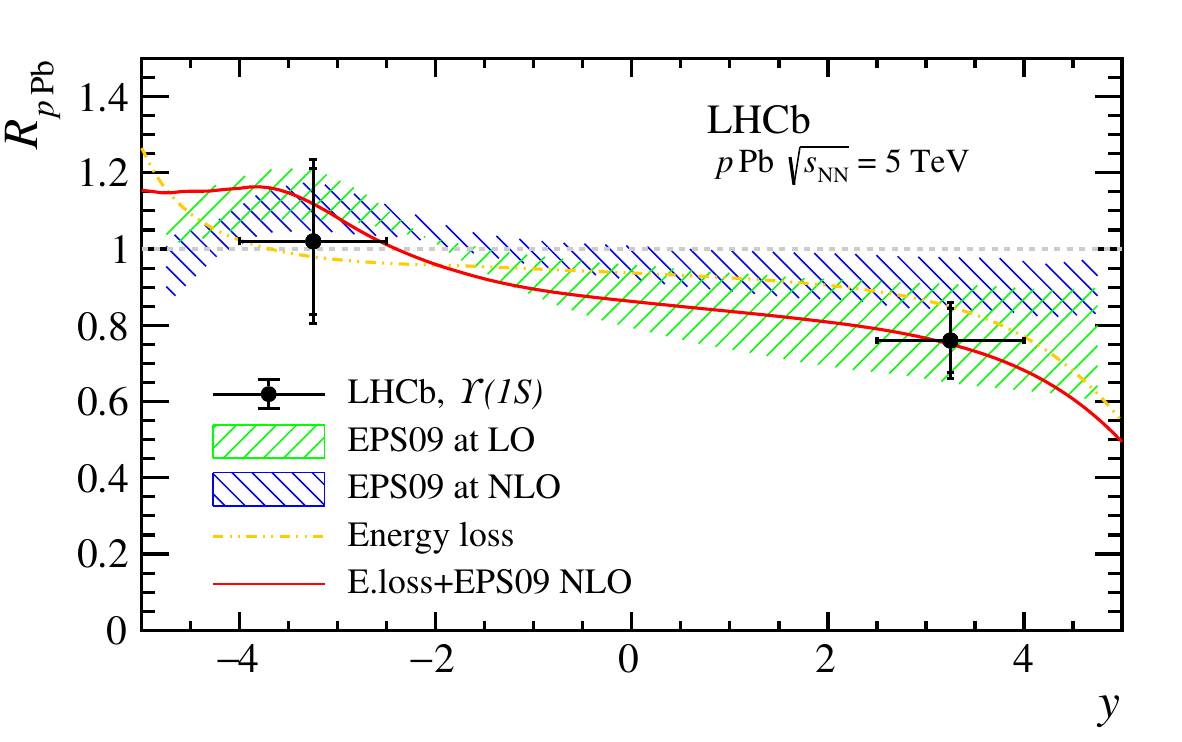}
    \caption{Nuclear modification factor $R_{p{\rm Pb}}$ as a function of $y$ for the $\OneS$ state, together with the theoretical predictions mentioned in the text. The inner error bars show the statistical uncertainties, while the outer ones show the statistical and systematic uncertainties added in quadrature.}
    \label{fig:Rppb}
\end{figure}

The measurement of the $\OneS$ cross-section times the $\OneS\to\mumu$ branching fraction in the same kinematic range from this analysis is given by 
\begin{equation*}
    \sigma(\OneS)\times\mathcal{B}(\emph{\OneS}\to\mumu)=1.34\pm0.02\pm0.05\nb,
\end{equation*}
where the first uncertainty is statistical and the second systematic.
While the two results are consistent, the direct measurement has an uncertainty that is a factor of two smaller.
The nuclear modification factor $R_{p{\rm Pb}}$ is updated based on the direct measurement and the results are
\begin{eqnarray}
    R_{p{\rm Pb}} =
    \begin{cases}
    1.02 \pm 0.19 \pm 0.10 &(-4.0<y<-2.5), \\
    0.76 \pm 0.08 \pm 0.05 &(2.5<y<4.0),
    \end{cases}
\end{eqnarray}
where the positive (negative) rapidity is for particles produced in the $p$ (Pb) beam direction.
Figure~\ref{fig:Rppb} compares the updated result with theoretical calculations  considering the EPS09 nuclear parton distribution function~\cite{Albacete:2013ei,Ferreiro:2011xy} or parton energy loss~\cite{Arleo:2012rs}, quoted in Ref.~\cite{LHCb-PAPER-2014-015}, and good agreements are found between data and these calculations.

\section{Conclusion}
The $\Upsilonres$ production cross-sections in proton-proton collisions
at a centre-of-mass energy of $\sqs=5\tev$ are studied
using a data sample with an integrated luminosity of \mbox{$9.13\pm0.18\invpb$}, 
collected by the \lhcb detector.
The double-differential cross-sections,
as function of the transverse momentum \pt and the rapidity $y$ of the $\Upsilonres$ mesons, are determined in the range $\pt<20\gevc$ and $2<y<4.5$ for the $\OneS$, $\TwoS$ and $\ThreeS$ states. The $\OneS$ cross-section as a function of $\pt$ is well described by the NRQCD calculation for $\pt>5\gevc$ with appropriate long distance matrix elements.
The ratios of  cross-sections of $\TwoS$ and $\ThreeS$ mesons with respect to $\OneS$ mesons are given in bins of \pt or $y$.
The ratios of cross-sections measured at different $pp$ collision energies are also determined.
The nuclear modification factor $R_{p{\rm Pb}}$ for $\OneS$ in proton-lead collisions at $\sqsnn = 5 \tev$ is updated using the $\OneS$ production in $pp$ data measured in this analysis. This result is consistent with the previous result, while improving  the precision by a factor of two.

\section*{Acknowledgements}
%
%
\noindent We express our gratitude to our colleagues in the CERN
accelerator departments for the excellent performance of the LHC. We
thank the technical and administrative staff at the LHCb
institutes.
We acknowledge support from CERN and from the national agencies:
CAPES, CNPq, FAPERJ and FINEP (Brazil); 
MOST and NSFC (China); 
CNRS/IN2P3 (France); 
BMBF, DFG and MPG (Germany); 
INFN (Italy); 
NWO (Netherlands); 
MNiSW and NCN (Poland); 
MEN/IFA (Romania); 
MICINN (Spain); 
SNSF and SER (Switzerland); 
NASU (Ukraine); 
STFC (United Kingdom); 
DOE NP and NSF (USA).
We acknowledge the computing resources that are provided by CERN, IN2P3
(France), KIT and DESY (Germany), INFN (Italy), SURF (Netherlands),
PIC (Spain), GridPP (United Kingdom), 
CSCS (Switzerland), IFIN-HH (Romania), CBPF (Brazil),
Polish WLCG  (Poland) and NERSC (USA).
We are indebted to the communities behind the multiple open-source
software packages on which we depend.
Individual groups or members have received support from
ARC and ARDC (Australia);
Minciencias (Colombia);
AvH Foundation (Germany);
EPLANET, Marie Sk\l{}odowska-Curie Actions, ERC and NextGenerationEU (European Union);
A*MIDEX, ANR, IPhU and Labex P2IO, and R\'{e}gion Auvergne-Rh\^{o}ne-Alpes (France);
Key Research Program of Frontier Sciences of CAS, CAS PIFI, CAS CCEPP, 
Fundamental Research Funds for the Central Universities, 
and Sci. \& Tech. Program of Guangzhou (China);
GVA, XuntaGal, GENCAT, Inditex, InTalent and Prog.~Atracci\'on Talento, CM (Spain);
SRC (Sweden);
the Leverhulme Trust, the Royal Society
 and UKRI (United Kingdom).

\newcommand{\xx}{\ensuremath{\kern 0.5em }}
\newcommand{\xxx}{\ensuremath{\kern 0.75em }}
\clearpage

\section*{Appendices}

\appendix
\setcounter{table}{0}
\setcounter{figure}{0}
\setcounter{equation}{0}

\section{Tabulated results}
\label{appendixA}
In the following, Tables~\ref{table1S}, \ref{table2S} and \ref{table3S} present the double-differential cross-sections times the $\Upsilonres\to\mumu$ branching fraction for $\OneS$, $\TwoS$ and $\ThreeS$ mesons, respectively.
Tables~\ref{table_pt} and \ref{table_y} present the cross-sections time the $\Upsilonres\to\mumu$ branching fraction as a function of $\pt$, integrated over $y$, or vice-versa.
Tables~\ref{tablenE_pt} and \ref{tablenE_y} present the ratios of cross-sections at $\sqs=13 \tev$ and $\sqs=5 \tev$ as a function of $\pt$ and $y$, respectively.

\label{sec:table}

\begin{sidewaystable}[!h]
	\centering
	\caption{Differential cross-section times the $\Upsilonres\to\mumu$ branching fraction (in unit of pb$/(\!\gevc)$) in bins of (\pt, $y$) for the $\OneS$ meson. The first uncertainties are statistical and the second are systematic.}
	
\begin{tabular}{l|lllll}
\hline \multicolumn{1}{c|}{\pt($\!\gevc$)} & \multicolumn{1}{c}{$2<y<2.5$} & \multicolumn{1}{c}{$2.5<y<3$} & \multicolumn{1}{c}{$3<y<3.5$} & \multicolumn{1}{c}{$3.5<y<4$} & \multicolumn{1}{c}{$4<y<4.5$}\\ \hline
0 $-$ 1&$50.6\pm 8.9\pm 2.3$&$53.6\pm 6.6\pm 2.2$&$53.6\pm 6.7\pm 2.3$&$45.7\pm 5.6\pm 2.1$&$18.2\pm 4.1\pm 1.3$\\
1 $-$ 2&$139.4\pm 14.8\pm 6.4$&$129.1\pm 10.2\pm 5.1$&$118.4\pm 9.4\pm 4.9$&$78.1\pm 7.5\pm 3.8$&$65.5\pm 7.9\pm 4.6$\\
2 $-$ 3&$163.2\pm 16.4\pm 7.8$&$173.3\pm 11.5\pm 6.9$&$141.5\pm 10.6\pm 5.8$&$97.0\pm 8.8\pm 4.8$&$53.8\pm 7.5\pm 3.9$\\
3 $-$ 4&$168.2\pm 16.5\pm 7.5$&$166.6\pm 11.7\pm 6.6$&$148.7\pm 10.5\pm 6.2$&$109.3\pm 9.7\pm 5.8$&$57.8\pm 8.0\pm 4.0$\\
4 $-$ 5&$150.2\pm 15.3\pm 6.5$&$149.5\pm 10.9\pm 6.0$&$116.4\pm 9.4\pm 4.8$&$73.7\pm 8.1\pm 4.1$&$47.5\pm 7.0\pm 3.3$\\
5 $-$ 6&$148.0\pm 15.0\pm 6.8$&$118.5\pm 9.8\pm 4.7$&$99.9\pm 8.4\pm 4.2$&$62.7\pm 7.7\pm 3.5$&$38.4\pm 6.8\pm 2.8$\\
6 $-$ 7&$101.6\pm 12.4\pm 4.7$&$89.4\pm 8.2\pm 3.6$&$70.7\pm 7.1\pm 3.0$&$53.9\pm 6.6\pm 3.0$&$19.9\pm 5.9\pm 1.5$\\
7 $-$ 8&$61.0\pm 10.2\pm 2.8$&$66.1\pm 7.2\pm 2.6$&$59.2\pm 6.0\pm 2.6$&\multirow{2}{*}{$24.5\pm 3.5\pm 1.4$}&\multirow{2}{*}{$16.0\pm 3.8\pm 1.2$}\\
8 $-$ 9&$42.1\pm 8.0\pm 1.9$&$49.4\pm 5.8\pm 2.0$&$34.3\pm 4.7\pm 1.5$&$ ~ $&  $ ~ $\\  
9 $-$ 10&$43.6\pm 8.3\pm 2.1$&$39.2\pm 5.2\pm 1.6$&$25.7\pm 4.0\pm 1.2$&\multirow{2}{*}{$12.8\pm 2.0\pm 0.8$}&\multirow{2}{*}{$4.4\pm 1.5\pm 0.3$}\\
10 $-$ 13&$21.0\pm 3.2\pm 1.0$&$22.1\pm 2.2\pm 0.9$&$11.4\pm 1.6\pm 0.5$&$ ~ $&  $ ~ $\\  
13 $-$ 20&$5.1\pm 1.1\pm 0.3$&$3.92\pm 0.65\pm 0.20$&$3.20\pm 0.57\pm 0.16$&$1.31\pm 0.45\pm 0.09$&$0.27\pm 0.30\pm 0.02$\\

\hline
\end{tabular}
	\label{table1S}
\end{sidewaystable}

\begin{sidewaystable}[!h]
	\centering
	\caption{Differential cross-section times the $\Upsilonres\to\mumu$ branching fraction (in unit of pb$/(\!\gevc)$) in bins of (\pt, $y$) for the $\TwoS$ meson. The first uncertainties are statistical and the second are systematic.}
	\begin{tabular}{l|lllll}
\hline \multicolumn{1}{c|}{\pt ($\!\gevc$)} & \multicolumn{1}{c}{$2<y<2.5$} & \multicolumn{1}{c}{$2.5<y<3$} & \multicolumn{1}{c}{$3<y<3.5$} & \multicolumn{1}{c}{$3.5<y<4$} & \multicolumn{1}{c}{$4<y<4.5$}\\ \hline
0 $-$ 1&$23.7\pm 6.5\pm 1.1$&$6.8\pm 3.2\pm 0.3$&$14.4\pm 4.0\pm 0.6$&$15.5\pm 3.4\pm 0.7$&$4.9\pm 2.6\pm 0.3$\\
1 $-$ 2&$38.6\pm 8.9\pm 1.8$&$34.4\pm 5.8\pm 1.4$&$22.0\pm 5.3\pm 1.0$&$26.2\pm 4.8\pm 1.3$&$5.0\pm 3.2\pm 0.4$\\
2 $-$ 3&$48.0\pm 9.6\pm 2.4$&$33.3\pm 6.0\pm 1.4$&$41.0\pm 6.4\pm 1.8$&$2.8\pm 4.1\pm 0.1$&$7.8\pm 3.5\pm 0.6$\\
3 $-$ 4&$54.4\pm 10.4\pm 2.5$&$30.9\pm 6.2\pm 1.3$&$31.9\pm 5.9\pm 1.4$&$18.4\pm 5.2\pm 1.0$&$7.1\pm 3.6\pm 0.5$\\
4 $-$ 5&$44.3\pm 9.3\pm 2.0$&$25.5\pm 5.7\pm 1.1$&$26.4\pm 5.4\pm 1.2$&$19.5\pm 5.1\pm 1.1$&$9.7\pm 3.8\pm 0.7$\\
5 $-$ 6&$34.1\pm 8.3\pm 1.6$&$23.0\pm 5.5\pm 1.0$&$33.0\pm 5.7\pm 1.5$&$13.3\pm 4.4\pm 0.8$&$3.20\pm 2.93\pm 0.24$\\
6 $-$ 7&$22.6\pm 7.7\pm 1.1$&$18.1\pm 4.5\pm 0.8$&$30.5\pm 5.3\pm 1.4$&$14.2\pm 4.3\pm 0.8$&$9.2\pm 3.9\pm 0.7$\\
7 $-$ 8&$18.4\pm 7.1\pm 0.9$&$22.1\pm 4.7\pm 0.9$&$14.7\pm 3.5\pm 0.7$&\multirow{2}{*}{$7.9\pm 2.4\pm 0.5$}&\multirow{2}{*}{$0.38\pm 2.90\pm 0.03$}\\
8 $-$ 9&$3.95\pm 4.10\pm 0.18$&$16.0\pm 3.7\pm 0.7$&$11.7\pm 3.2\pm 0.5$&$ ~ $&  $ ~ $\\  
9 $-$ 10&$8.8\pm 4.7\pm 0.4$&$12.9\pm 3.8\pm 0.5$&$9.4\pm 2.4\pm 0.4$&\multirow{2}{*}{$3.49\pm 1.08\pm 0.22$}&\multirow{2}{*}{$1.68\pm 1.00\pm 0.12$}\\
10 $-$ 13&$9.4\pm 2.6\pm 0.5$&$6.4\pm 1.4\pm 0.3$&$4.40\pm 0.98\pm 0.21$&$ ~ $&  $ ~ $\\  
13 $-$ 20&$3.01\pm 0.86\pm 0.16$&$1.44\pm 0.53\pm 0.07$&$1.09\pm 0.37\pm 0.05$&$0.41\pm 0.37\pm 0.03$&$0.16\pm 0.21\pm 0.02$\\
\hline
\end{tabular}
	\label{table2S}
\end{sidewaystable}

\begin{sidewaystable}[!h]
	\centering
	\caption{Differential cross-section times the $\Upsilonres\to\mumu$ branching fraction (in unit of pb$/(\!\gevc)$) in bins of (\pt, $y$) for the $\ThreeS$ meson. The first uncertainties are statistical and the second are systematic.}
	\begin{tabular}{l|lllll}
\hline \multicolumn{1}{c|}{\pt($\!\gevc$)} & \multicolumn{1}{c}{$2<y<2.5$} & \multicolumn{1}{c}{$2.5<y<3$} & \multicolumn{1}{c}{$3<y<3.5$} & \multicolumn{1}{c}{$3.5<y<4$} & \multicolumn{1}{c}{$4<y<4.5$}\\ \hline
0 $-$ 1&  $4.31\pm 3.50\pm 0.21$&  $0.68\pm 2.34\pm 0.03$&  $5.63\pm 3.23\pm 0.26$&  $5.74\pm 2.41\pm 0.28$&  $1.72\pm 1.66\pm 0.13$\\  
1 $-$ 2&  $10.53\pm 6.6\pm 0.51$&  $10.09\pm 4.33\pm 0.44$&  $14.68\pm 4.45\pm 0.68$&  $4.56\pm 3.01\pm 0.24$&  $2.42\pm 2.62\pm 0.17$\\  
2 $-$ 3&  $21.82\pm 7.16\pm 1.10$&  $13.92\pm 4.83\pm 0.61$&  $7.44\pm 4.66\pm 0.34$&  $7.77\pm 4.15\pm 0.43$&  $6.43\pm 3.35\pm 0.48$\\  
3 $-$ 4&  $7.33\pm 6.31\pm 0.35$&  $17.17\pm 5.19\pm 0.75$&  $21.45\pm 5.12\pm 0.99$&  $9.15\pm 4.53\pm 0.54$&  $8.10\pm 3.86\pm 0.59$\\  
4 $-$ 5&  $15.12\pm 6.33\pm 0.7$&  $21.59\pm 5.18\pm 0.95$&  $20.60\pm 4.86\pm 0.96$&  $10.02\pm 4.44\pm 0.59$&  $1.11\pm 2.87\pm 0.08$\\  
5 $-$ 6&  $6.81\pm 5.67\pm 0.33$&  $11.17\pm 4.42\pm 0.49$&  $18.19\pm 4.91\pm 0.85$&  $9.11\pm 3.98\pm 0.54$&  $1.14\pm 3.2\pm 0.09$\\  
6 $-$ 7&  $10.28\pm 5.0\pm 0.51$&  $8.07\pm 3.65\pm 0.35$&  $11.79\pm 4.30\pm 0.55$&  $9.5\pm 3.39\pm 0.56$&  $5.76\pm 3.22\pm 0.43$\\  
7 $-$ 8&  $11.90\pm 5.60\pm 0.58$&  $6.12\pm 3.45\pm 0.27$&  $6.87\pm 2.88\pm 0.32$&  \multirow{2}{*}{$3.65\pm 2.19\pm 0.23$}&  \multirow{2}{*}{$0.00\pm 2.21\pm 0.00$}\\  
8 $-$ 9&  $0.52\pm 7.65\pm 0.03$&  $10.79\pm 3.33\pm 0.47$&  $2.8\pm 2.19\pm 0.14$&  $ ~ $&  $ ~ $\\  
9 $-$ 10&  $12.20\pm 5.10\pm 0.61$&  $3.28\pm 2.56\pm 0.14$&  $5.02\pm 1.96\pm 0.25$&  \multirow{2}{*}{$2.02\pm 1.03\pm 0.13$}&  \multirow{2}{*}{$1.43\pm 1.12\pm 0.10$}\\  
10 $-$ 13&  $4.76\pm 2.02\pm 0.24$&  $3.18\pm 1.19\pm 0.14$&  $3.47\pm 0.91\pm 0.17$&  $ ~ $&  $ ~ $\\  
13 $-$ 20&  $1.24\pm 0.69\pm 0.07$&  $1.64\pm 0.48\pm 0.09$&  $0.85\pm 0.34\pm 0.04$&  $0.07\pm 0.38\pm 0.00$&  $0.23\pm 0.32\pm 0.02$\\
\end{tabular}
	\label{table3S}
\end{sidewaystable}

\begin{table}[!h]
	\centering
	\caption{Differential cross-section times the $\Upsilonres\to\mumu$ branching fraction (in unit of pb$/(\!\gevc)$) of the $\OneS$, $\TwoS$ and $\ThreeS$ mesons as a function of $\pt$ for $y$ integrated from 2.0 to 4.5.}
	\begin{tabular}{l|lll}
\hline \multicolumn{1}{c|}{\pt($\!\gevc$)} & \multicolumn{1}{c}{$\OneS$} & \multicolumn{1}{c}{$\TwoS$} & \multicolumn{1}{c}{$\ThreeS$} \\ \hline
\phz 0 $-$ 1 & $111\phz~\pm\phz8\phz~\pm\phz5$      & $32.6\pm4.7\pm1.5$        &$\phz9.0\pm3.0\pm0.5$\\
\phz 1 $-$ 2 & $265\phz~\pm12\phz~\pm12$            & $63.1\pm6.6\pm2.9$        &$21.1\pm5.0\pm1.0$\\
\phz 2 $-$ 3 & $314\phz~\pm13\phz~\pm15$            & $66.5\pm7.0\pm3.1$        &$28.7\pm5.6\pm1.5$\\
\phz 3 $-$ 4 & $325\phz~\pm13\phz~\pm15$            & $71.3\pm7.5\pm3.3$        &$31.6\pm5.7\pm1.6$\\
\phz 4 $-$ 5 & $269\phz~\pm12\phz~\pm12$            & $62.7\pm6.9\pm3.0$        &$34.2\pm5.4\pm1.6$\\
\phz 5 $-$ 6 & $234\phz~\pm11\phz~\pm11$            & $53.3\pm6.3\pm2.5$        &$23.2\pm5.1\pm1.2$\\
\phz 6 $-$ 7 & $162\phz~\pm10\phz~\pm\phz8$         & $44.9\pm5.7\pm2.2$        &$22.7\pm4.4\pm1.2$\\
\phz 7 $-$ 9 & $\phz98\phz~\pm\phz7\phz~\pm\phz5$   & $25.8\pm4.4\pm1.2$        &$11.6\pm4.3\pm0.6$\\
\phz 9 $-$ 13& $\phz42.6\pm\phz3.4\pm\phz2.1$       & $14.1\pm2.2\pm0.7$        &$\phz8.6\pm2.0\pm0.5$\\
13 $-$ 20    & $\phz\phz 6.9\pm\phz0.8\pm\phz0.4$   & $\phz3.1\pm0.6\pm0.2$     &$\phz2.0\pm0.5\pm0.1$\\
\hline
\end{tabular}

	\label{table_pt}
\end{table}

\begin{table}[!h]
    \centering
	\caption{Differential cross-section times the $\Upsilonres\to\mumu$ branching fraction (in unit of pb) of $\OneS$, $\TwoS$ and $\ThreeS$ mesons as a function $y$ for $\pt$ integrated from 0 to 20 $\gevc$.}
	\begin{tabular}{l|lll}
\hline \multicolumn{1}{c|}{$y$} & \multicolumn{1}{c}{$\OneS$} & \multicolumn{1}{c}{$\TwoS$} & \multicolumn{1}{c}{$\ThreeS$} \\ \hline
$2.0<y<2.5$   &$1167\pm 43\pm 45$     &$346\pm 27\pm 14$      &$124\phz~\pm21\phz~\pm5$\\
$2.5<y<3.0$   &$1129\pm 30\pm 41$     &$252\pm 17\pm 10$      &$124\phz~\pm14\phz~\pm5$\\
$3.0<y<3.5$   &$\phz925\pm 26\pm 35$  &$256\pm 16\pm 10$      &$131\phz~\pm13\phz~\pm5$\\
$3.5<y<4.0$   &$\phz630\pm 24\pm 26$  &$143\pm 14\pm 6$       &$\phz71.7\pm12.0\pm3.2$\\
$4.0<y<4.5$   &$\phz353\pm 21\pm 19$  &$\phz55\pm 12\pm 3$    &$\phz34.0\pm10.4\pm1.8$\\
\hline
\end{tabular}
	\label{table_y}
\end{table}

\begin{table}[!h]
	\centering
	\caption{Cross-section ratios between $\sqrt{s}=13\tev$ and $5\tev$ of $\OneS$, $\TwoS$ and $\ThreeS$ mesons as a function of $\pt$ for $y$ integrated from 2.0 to 4.5.}
	\begin{tabular}{l|lll}
\hline \multicolumn{1}{c|}{\pt($\!\gevc$)} & \multicolumn{1}{c}{$\OneS$} & \multicolumn{1}{c}{$\TwoS$} & \multicolumn{1}{c}{$\ThreeS$} \\ \hline
\phz 0 $-$ 1 &$1.74\pm 0.12\pm 0.07$&$1.17\pm 0.17\pm 0.05$&$1.85\pm 0.62\pm 0.09$\\
\phz 1 $-$ 2 &$1.87\pm 0.08\pm 0.08$&$1.59\pm 0.17\pm 0.07$&$2.00\pm 0.47\pm 0.09$\\
\phz 2 $-$ 3 &$2.04\pm 0.08\pm 0.09$&$1.96\pm 0.21\pm 0.09$&$2.03\pm 0.39\pm 0.10$\\
\phz 3 $-$ 4 &$2.03\pm 0.08\pm 0.09$&$2.00\pm 0.21\pm 0.09$&$2.03\pm 0.36\pm 0.10$\\
\phz 4 $-$ 5 &$2.25\pm 0.10\pm 0.10$&$2.19\pm 0.24\pm 0.10$&$1.85\pm 0.29\pm 0.08$\\
\phz 5 $-$ 6 &$2.16\pm 0.10\pm 0.10$&$2.30\pm 0.27\pm 0.10$&$2.64\pm 0.57\pm 0.13$\\
\phz 6 $-$ 7 &$2.44\pm 0.15\pm 0.11$&$2.24\pm 0.29\pm 0.11$&$2.38\pm 0.46\pm 0.12$\\
\phz 7 $-$ 9 &$2.88\pm 0.15\pm 0.14$&$3.04\pm 0.40\pm 0.15$&$3.65\pm 1.01\pm 0.19$\\
\phz 9 $-$ 13&$2.99\pm 0.17\pm 0.19$&$2.87\pm 0.33\pm 0.16$&$2.72\pm 0.45\pm 0.17$\\
13 $-$ 20    &$4.32\pm 0.48\pm 0.24$&$3.76\pm 0.71\pm 0.21$&$3.61\pm 0.93\pm 0.19$\\
\hline
\end{tabular}

	\label{tablenE_pt}
\end{table}

\begin{table}[!h]
    \centering
	\caption{Cross-section ratios between $\sqrt{s}=13\tev$ and $5\tev$ of $\OneS$, $\TwoS$ and $\ThreeS$ mesons as a function of $y$ for $\pt$ integrated from 0 to 20 $\gevc$.}
	\begin{tabular}{l|lll}
\hline \multicolumn{1}{c|}{$y$} & \multicolumn{1}{c}{$\OneS$} & \multicolumn{1}{c}{$\TwoS$} & \multicolumn{1}{c}{$\ThreeS$} \\ \hline
$2.0<y<2.5$&$2.13\pm 0.08\pm 0.05$&$1.79\pm 0.14\pm 0.05$&$2.49\pm 0.41\pm 0.07$\\
$2.5<y<3.0$&$2.04\pm 0.05\pm 0.05$&$2.23\pm 0.15\pm 0.05$&$2.32\pm 0.26\pm 0.06$\\
$3.0<y<3.5$&$2.19\pm 0.06\pm 0.05$&$1.94\pm 0.12\pm 0.05$&$1.89\pm 0.19\pm 0.05$\\
$3.5<y<4.0$&$2.66\pm 0.10\pm 0.07$&$2.86\pm 0.28\pm 0.08$&$2.78\pm 0.46\pm 0.08$\\
$4.0<y<4.5$&$3.11\pm 0.18\pm 0.11$&$4.80\pm 1.00\pm 0.16$&$3.94\pm 1.21\pm 0.13$\\
\hline
\end{tabular}

	\label{tablenE_y}
\end{table}

\clearpage

\section{Dependence of cross-sections on the polarisation}
\label{appendixB}

The angular distribution of $\Upsilonres \to \mumu$ decays is described by
\begin{equation}
    \frac{\deriv^2N}{\deriv\cos\theta\deriv\phi}\propto 1+\lambda_{\theta}\cos^2\theta+\lambda_{\theta\phi}\sin2\theta\cos\phi+\lambda_{\phi}\sin^2\theta\cos2\phi,
\end{equation}
where $\theta$ and $\phi$ are the polar and azimuthal angles of the $\mu^+$ momentum in the rest frame of the $\Upsilonres$  meson for a given polarisation coordinate system, and $\lambda_{\theta}$, $\lambda_{\theta\phi}$ and $\lambda_{\phi}$ are the three polarisation parameters~\cite{Faccioli:2010kd}.
Zero polarisation implies $\lambda_{\theta}=\lambda_{\theta\phi}=\lambda_{\phi}=0$.
The detection efficiency of the \Upsilonres mesons is a function of the polarisation, especially of $\lambda_{\theta}$.

The $\Upsilonres(nS)$ mesons are assumed to be unpolarised in this measurement, which is also the case for the simulated signal samples used to determine the efficiencies.
Such an assumption is supported by the measurements of the $\Upsilonres$ polarization by LHCb in $pp$ collisions at $\sqrt{s}=7 \tev$ and $8 \tev$ \cite{LHCb-PAPER-2017-028} 
in the same kinematic range as in this analysis, and  by the CMS experiment \cite{CMS:2012bpf} for $\sqrt{s}=8\tev$.

To evaluate the change of results assuming  non-zero polarisation,
the angular distributions of muons in the $\Upsilonres$ rest frame in the simulated signal samples are reweighted, and the total efficiencies are recomputed.
The relative changes of the double-differential cross-sections for a polarisation of $\lambda_{\theta}=0.1$~\cite{LHCb-PAPER-2017-028}  in the helicity frame compared to zero polarisation in each (\pt,$y$) interval are given in Tables~\ref{table_0.1pol_1S}, \ref{table_0.1pol_2S} and \ref{table_0.1pol_3S} for $\OneS$, $\TwoS$ and $\ThreeS$ mesons, respectively. In addition, the relative changes of the double-differential cross-sections for a polarisation of $\lambda_{\theta}=+1\,(-1)$ are also evaluated, and are shown in Tables~\ref{table_1pol_1S}(\ref{table_-1pol_1S}), \ref{table_1pol_2S}(\ref{table_-1pol_2S}) and \ref{table_1pol_3S}(\ref{table_-1pol_3S}) for $\OneS$, $\TwoS$ and $\ThreeS$ mesons, respectively. The helicity frame uses the $\Upsilonres$ momentum in the laboratory frame as the spin quantization axis.

\begin{table}[h]
	\centering
	\caption{Relative changes of double-differential cross-sections of $\Upsilonres(1S)$ (in \%), for a polarisation of $\lambda_{\theta}=0.1$ rather than zero, in (\pt,$y$) intervals.}
        \label{table_0.1pol_1S}
	\begin{tabular}{l|lllll}
\hline \pt($\!\gevc$) & $2<y<2.5$ & $2.5<y<3$ & $3<y<3.5$ & $3.5<y<4$ & $4<y<4.5$\\ \hline

0 $-$ 1&  $2.87\pm0.37$  &  $1.83\pm0.13$  &  $1.01\pm0.07$  &  $\phz~1.26\pm0.13$  &  $\phz~2.06\pm0.38$    \\  
1 $-$ 2&  $2.75\pm0.21$  &  $1.83\pm0.08$  &  $1.06\pm0.05$  &  $\phz~1.15\pm0.08$  &  $\phz~1.93\pm0.23$    \\  
2 $-$ 3&  $2.72\pm0.18$  &  $1.78\pm0.07$  &  $1.10\pm0.05$  &  $\phz~1.03\pm0.08$  &  $\phz~1.62\pm0.20$    \\  
3 $-$ 4&  $2.60\pm0.17$  &  $1.75\pm0.07$  &  $1.03\pm0.05$  &  $\phz~0.84\pm0.08$  &  $\phz~1.16\pm0.20$    \\  
4 $-$ 5&  $2.45\pm0.17$  &  $1.68\pm0.07$  &  $0.94\pm0.06$  &  $\phz~0.66\pm0.10$  &  $\phz~0.78\pm0.22$    \\  
5 $-$ 6&  $2.30\pm0.19$  &  $1.63\pm0.08$  &  $0.78\pm0.07$  &  $\phz~0.41\pm0.11$  &  $\phz~0.39\pm0.25$    \\  
6 $-$ 7&  $2.16\pm0.21$  &  $1.40\pm0.09$  &  $0.61\pm0.08$  &  $\phz~0.19\pm0.14$  &  $\phz~0.16\pm0.30$    \\  
7 $-$ 8&  $2.02\pm0.23$  &  $1.34\pm0.10$  &  $0.47\pm0.10$  &  \multirow{2}{*}{$-0.05\pm0.12$}  &  \multirow{2}{*}{$-0.24\pm0.27$}    \\  
8 $-$ 9&  $1.84\pm0.25$  &  $1.25\pm0.12$  &  $0.40\pm0.11$  &  $ ~ $&  $ ~ $  \\  
9 $-$ 10&  $1.75\pm0.28$  &  $1.14\pm0.13$  &  $0.34\pm0.14$  &  \multirow{2}{*}{$-0.17\pm0.14$}  &  \multirow{2}{*}{$-0.60\pm0.32$}    \\  
10 $-$ 13&  $1.48\pm0.19$  &  $1.07\pm0.09$  &  $0.28\pm0.11$  &  $ ~ $&  $ ~ $  \\  
13 $-$ 20&  $1.34\pm0.22$  &  $0.85\pm0.11$  &  $0.28\pm0.13$  &  $-0.12\pm0.23$  &  $-0.56\pm0.59$    \\  

\hline
\end{tabular}

\end{table}

\begin{table}[h]
	\centering
	\caption{Relative changes of cross-sections of $\Upsilonres(2S)$ (in \%), for a polarisation of $\lambda_{\theta}=0.1$ rather than zero, in (\pt,$y$) intervals.}
        \label{table_0.1pol_2S}
	\begin{tabular}{l|lllll}
\hline \pt($\!\gevc$) & $2<y<2.5$ & $2.5<y<3$ & $3<y<3.5$ & $3.5<y<4$ & $4<y<4.5$\\ \hline

0 $-$ 1&  $2.83\pm0.33$  &  $1.88\pm0.14$  &  $1.03\pm0.09$  &  $\phz~1.29\pm0.15$  &  $\phz~2.18\pm0.40$    \\  
1 $-$ 2&  $2.78\pm0.20$  &  $1.83\pm0.08$  &  $1.06\pm0.05$  &  $\phz~1.19\pm0.09$  &  $\phz~1.96\pm0.24$    \\  
2 $-$ 3&  $2.73\pm0.17$  &  $1.78\pm0.07$  &  $1.06\pm0.05$  &  $\phz~1.05\pm0.08$  &  $\phz~1.67\pm0.21$    \\  
3 $-$ 4&  $2.61\pm0.16$  &  $1.74\pm0.07$  &  $1.02\pm0.05$  &  $\phz~0.89\pm0.08$  &  $\phz~1.23\pm0.20$    \\  
4 $-$ 5&  $2.46\pm0.17$  &  $1.68\pm0.07$  &  $0.94\pm0.06$  &  $\phz~0.67\pm0.10$  &  $\phz~0.84\pm0.22$    \\  
5 $-$ 6&  $2.33\pm0.18$  &  $1.64\pm0.08$  &  $0.83\pm0.07$  &  $\phz~0.48\pm0.11$  &  $\phz~0.51\pm0.26$    \\  
6 $-$ 7&  $2.18\pm0.20$  &  $1.46\pm0.08$  &  $0.63\pm0.07$  &  $\phz~0.24\pm0.14$  &  $\phz~0.21\pm0.30$    \\  
7 $-$ 8&  $2.08\pm0.23$  &  $1.41\pm0.10$  &  $0.50\pm0.09$  &  \multirow{2}{*}{$\phz~0.02\pm0.12$}  &  \multirow{2}{*}{$-0.17\pm0.29$}    \\  
8 $-$ 9&  $1.88\pm0.25$  &  $1.23\pm0.10$  &  $0.49\pm0.10$  &  $ ~ $&  $ ~ $ \\  
9 $-$ 10&  $1.84\pm0.29$  &  $1.16\pm0.12$  &  $0.31\pm0.12$  &  \multirow{2}{*}{$-0.18\pm0.15$}  &  \multirow{2}{*}{$-0.45\pm0.33$}    \\  
10 $-$ 13&  $1.59\pm0.20$  &  $1.11\pm0.08$  &  $0.27\pm0.10$  &  $ ~ $&  $ ~ $  \\  
13 $-$ 20&  $1.34\pm0.20$  &  $0.89\pm0.08$  &  $0.28\pm0.11$  &  $-0.15\pm0.22$  &  $-0.57\pm0.65$    \\  

\hline
\end{tabular}

\end{table}

\begin{table}[h]
	\centering
	\caption{Relative changes of cross-sections of $\Upsilonres(3S)$ (in \%), for a polarisation of $\lambda_{\theta}=0.1$ rather than zero, in (\pt,$y$) intervals}
        \label{table_0.1pol_3S}
	\begin{tabular}{l|lllll}
\hline \pt($\!\gevc$) & $2<y<2.5$ & $2.5<y<3$ & $3<y<3.5$ & $3.5<y<4$ & $4<y<4.5$\\ \hline

0 $-$ 1&  $2.81\pm0.32$  &  $1.87\pm0.13$  &  $1.04\pm0.08$  &  $\phz~1.22\pm0.14$  &  $\phz~2.27\pm0.44$    \\  
1 $-$ 2&  $2.77\pm0.20$  &  $1.84\pm0.08$  &  $1.06\pm0.05$  &  $\phz~1.20\pm0.09$  &  $\phz~1.97\pm0.24$    \\  
2 $-$ 3&  $2.72\pm0.17$  &  $1.81\pm0.07$  &  $1.07\pm0.05$  &  $\phz~1.09\pm0.08$  &  $\phz~1.64\pm0.21$    \\  
3 $-$ 4&  $2.62\pm0.16$  &  $1.77\pm0.07$  &  $1.02\pm0.05$  &  $\phz~0.91\pm0.08$  &  $\phz~1.28\pm0.20$    \\  
4 $-$ 5&  $2.54\pm0.18$  &  $1.70\pm0.07$  &  $0.95\pm0.06$  &  $\phz~0.72\pm0.10$  &  $\phz~0.91\pm0.22$    \\  
5 $-$ 6&  $2.33\pm0.18$  &  $1.65\pm0.08$  &  $0.84\pm0.07$  &  $\phz~0.51\pm0.11$  &  $\phz~0.56\pm0.27$    \\  
6 $-$ 7&  $2.21\pm0.21$  &  $1.51\pm0.09$  &  $0.69\pm0.08$  &  $\phz~0.29\pm0.14$  &  $\phz~0.29\pm0.31$    \\  
7 $-$ 8&  $2.11\pm0.24$  &  $1.43\pm0.10$  &  $0.53\pm0.09$  &  \multirow{2}{*}{$\phz~0.09\pm0.12$}  &  \multirow{2}{*}{$-0.06\pm0.28$}    \\  
8 $-$ 9&  $1.94\pm0.26$  &  $1.23\pm0.10$  &  $0.44\pm0.10$  &  $ ~ $&  $ ~ $  \\  
9 $-$ 10&  $1.74\pm0.28$  &  $1.22\pm0.11$  &  $0.42\pm0.10$  &  \multirow{2}{*}{$-0.11\pm0.14$}  &  \multirow{2}{*}{$-0.52\pm0.33$}    \\  
10 $-$ 13&  $1.60\pm0.19$  &  $1.05\pm0.07$  &  $0.31\pm0.08$  &  $ ~ $&  $ ~ $  \\  
13 $-$ 20&  $1.42\pm0.21$  &  $0.87\pm0.07$  &  $0.34\pm0.08$  &  $-0.11\pm0.18$  &  $-0.55\pm0.62$    \\  

\hline
\end{tabular}

\end{table}

\begin{table}[h]
	\centering
	\caption{Relative changes of cross-sections of $\Upsilonres(1S)$ (in \%), for a polarisation of $\lambda_{\theta}=1$ rather than zero, in (\pt,$y$) intervals}
        \label{table_1pol_1S}
	\begin{tabular}{l|lllll}
\hline \pt($\!\gevc$) & $2<y<2.5$ & $2.5<y<3$ & $3<y<3.5$ & $3.5<y<4$ & $4<y<4.5$\\ \hline

0 $-$ 1&  $27.5\pm1.5$  &  $16.2\pm0.5$  &  $8.4\pm0.3$  &  $~10.7\pm0.5$  &  $~18.8\pm1.4$    \\  
1 $-$ 2&  $26.3\pm0.8$  &  $16.2\pm0.3$  &  $8.8\pm0.2$  &  $\phz~9.7\pm0.3$  &  $~17.2\pm0.9$    \\  
2 $-$ 3&  $25.8\pm0.7$  &  $15.7\pm0.3$  &  $9.2\pm0.2$  &  $\phz~8.6\pm0.3$  &  $~14.2\pm0.7$    \\  
3 $-$ 4&  $24.4\pm0.7$  &  $15.4\pm0.3$  &  $8.6\pm0.2$  &  $\phz~6.9\pm0.3$  &  $\phz~9.8\pm0.6$    \\  
4 $-$ 5&  $22.7\pm0.7$  &  $14.7\pm0.3$  &  $7.8\pm0.2$  &  $\phz~5.3\pm0.3$  &  $\phz~6.4\pm0.7$    \\  
5 $-$ 6&  $21.1\pm0.7$  &  $14.2\pm0.3$  &  $6.4\pm0.2$  &  $\phz~3.3\pm0.3$  &  $\phz~3.1\pm0.7$    \\  
6 $-$ 7&  $19.5\pm0.8$  &  $12.1\pm0.3$  &  $4.9\pm0.3$  &  $\phz~1.5\pm0.4$  &  $\phz~1.3\pm0.9$    \\  
7 $-$ 8&  $18.0\pm0.8$  &  $11.4\pm0.4$  &  $3.8\pm0.3$  &  \multirow{2}{*}{$-0.4\pm0.3$}  &  \multirow{2}{*}{$-1.8\pm0.7$}    \\  
8 $-$ 9&  $16.3\pm0.9$  &  $10.6\pm0.4$  &  $3.2\pm0.4$  &  $ ~ $&  $ ~ $  \\  
9 $-$ 10&  $15.4\pm1.0$  &  $\phz9.6\pm0.5$  &  $2.7\pm0.4$  &  \multirow{2}{*}{$-1.3\pm0.4$}  &  \multirow{2}{*}{$-4.5\pm0.8$}    \\  
10 $-$ 13&  $12.8\pm0.7$  &  $\phz8.9\pm0.3$  &  $2.2\pm0.3$  &  $ ~ $&  $ ~ $  \\  
13 $-$ 20&  $11.4\pm0.8$  &  $\phz7.0\pm0.4$  &  $2.2\pm0.4$  &  $-0.9\pm0.6$  &  $-4.2\pm1.5$    \\   

\hline
\end{tabular}
\end{table}

\begin{table}[h]
	\centering
	\caption{Relative changes of cross-sections of $\Upsilonres(2S)$ (in \%), for a polarisation of $\lambda_{\theta}=1$ rather than zero, in (\pt,$y$) intervals.}
        \label{table_1pol_2S}
	\begin{tabular}{l|lllll}
\hline \pt($\!\gevc$) & $2<y<2.5$ & $2.5<y<3$ & $3<y<3.5$ & $3.5<y<4$ & $4<y<4.5$\\ \hline

0 $-$ 1&  $27.2\pm1.3$  &  $16.7\pm0.5$  &  $8.5\pm0.3$  &  $~11.0\pm0.5$  &  $~19.8\pm1.5$    \\  
1 $-$ 2&  $26.6\pm0.8$  &  $16.2\pm0.3$  &  $8.8\pm0.2$  &  $~10.0\pm0.3$  &  $~17.6\pm0.9$    \\  
2 $-$ 3&  $25.9\pm0.7$  &  $15.7\pm0.3$  &  $8.9\pm0.2$  &  $\phz~8.8\pm0.3$  &  $~14.6\pm0.7$    \\  
3 $-$ 4&  $24.5\pm0.6$  &  $15.3\pm0.3$  &  $8.5\pm0.2$  &  $\phz~7.3\pm0.3$  &  $~10.4\pm0.7$    \\  
4 $-$ 5&  $22.9\pm0.6$  &  $14.7\pm0.3$  &  $7.8\pm0.2$  &  $\phz~5.5\pm0.3$  &  $\phz~6.9\pm0.7$    \\  
5 $-$ 6&  $21.5\pm0.7$  &  $14.3\pm0.3$  &  $6.8\pm0.2$  &  $\phz~3.8\pm0.3$  &  $\phz~4.1\pm0.8$    \\  
6 $-$ 7&  $19.8\pm0.8$  &  $12.5\pm0.3$  &  $5.1\pm0.3$  &  $\phz~1.9\pm0.4$  &  $\phz~1.7\pm0.9$    \\  
7 $-$ 8&  $18.6\pm0.9$  &  $12.1\pm0.4$  &  $4.0\pm0.3$  &  \multirow{2}{*}{$\phz~0.2\pm0.4$}  &  \multirow{2}{*}{$-1.3\pm0.8$}    \\  
8 $-$ 9&  $16.7\pm0.9$  &  $10.4\pm0.4$  &  $3.9\pm0.3$  &  $ ~ $&  $ ~ $ \\  
9 $-$ 10&  $16.1\pm1.1$  &  $\phz9.8\pm0.5$  &  $2.5\pm0.4$  &  \multirow{2}{*}{$-1.4\pm0.4$}  &  \multirow{2}{*}{$-3.4\pm0.9$}    \\  
10 $-$ 13&  $13.7\pm0.7$  &  $\phz9.2\pm0.3$  &  $2.1\pm0.3$  &  $ ~ $&  $ ~ $  \\  
13 $-$ 20&  $11.5\pm0.7$  &  $\phz7.3\pm0.3$  &  $2.2\pm0.4$  &  $-1.2\pm0.6$  &  $-4.3\pm1.6$    \\  

\hline
\end{tabular}

\end{table}

\begin{table}[h]
	\centering
	\caption{Relative changes of cross-sections of $\Upsilonres(3S)$ (in \%), for a polarisation of $\lambda_{\theta}=1$ rather than zero, in (\pt,$y$) intervals.}
        \label{table_1pol_3S}
	\begin{tabular}{l|lllll}
\hline \pt($\!\gevc$) & $2<y<2.5$ & $2.5<y<3$ & $3<y<3.5$ & $3.5<y<4$ & $4<y<4.5$\\ \hline

0 $-$ 1&  $27.0\pm1.3$  &  $16.6\pm0.5$  &  $8.7\pm0.3$  &  $~10.3\pm0.5$  &  $~20.6\pm1.6$    \\  
1 $-$ 2&  $26.5\pm0.8$  &  $16.3\pm0.3$  &  $8.9\pm0.2$  &  $~10.1\pm0.3$  &  $~17.6\pm0.9$    \\  
2 $-$ 3&  $25.9\pm0.7$  &  $15.9\pm0.3$  &  $8.9\pm0.2$  &  $\phz~9.2\pm0.3$  &  $~14.4\pm0.7$    \\  
3 $-$ 4&  $24.6\pm0.6$  &  $15.5\pm0.3$  &  $8.5\pm0.2$  &  $\phz~7.5\pm0.3$  &  $~10.9\pm0.7$    \\  
4 $-$ 5&  $23.6\pm0.7$  &  $14.9\pm0.3$  &  $7.9\pm0.2$  &  $\phz~5.9\pm0.3$  &  $\phz~7.5\pm0.7$    \\  
5 $-$ 6&  $21.5\pm0.7$  &  $14.4\pm0.3$  &  $6.9\pm0.2$  &  $\phz~4.1\pm0.4$  &  $\phz~4.5\pm0.8$    \\  
6 $-$ 7&  $20.2\pm0.8$  &  $13.0\pm0.4$  &  $5.6\pm0.3$  &  $\phz~2.3\pm0.4$  &  $\phz~2.3\pm0.9$    \\  
7 $-$ 8&  $18.9\pm0.9$  &  $12.3\pm0.4$  &  $4.3\pm0.3$  &  \multirow{2}{*}{$\phz~0.7\pm0.4$}  &  \multirow{2}{*}{$-0.5\pm0.8$}    \\  
8 $-$ 9&  $17.3\pm1.0$  &  $10.4\pm0.4$  &  $3.5\pm0.3$  &  $ ~ $&  $ ~ $ \\  
9 $-$ 10&  $15.3\pm1.0$  &  $10.3\pm0.5$  &  $3.4\pm0.4$  &  \multirow{2}{*}{$-0.8\pm0.4$}  &  \multirow{2}{*}{$-3.8\pm0.9$}    \\  
10 $-$ 13&  $14.0\pm0.7$  &  $\phz8.8\pm0.3$  &  $2.4\pm0.3$  &  $ ~ $&  $ ~ $ \\  
13 $-$ 20&  $12.1\pm0.7$  &  $\phz7.2\pm0.3$  &  $2.7\pm0.3$  &  $-0.8\pm0.5$  &  $-4.2\pm1.6$   \\  

\hline
\end{tabular}

\end{table}

\begin{table}[h]
	\centering
	\caption{Relative changes of cross-sections of $\Upsilonres(1S)$ (in \%), for a polarisation of $\lambda_{\theta}=-1$ rather than zero, in (\pt,$y$) intervals.}
        \label{table_-1pol_1S}
	\begin{tabular}{l|lllll}
\hline \pt($\!\gevc$) & $2<y<2.5$ & $2.5<y<3$ & $3<y<3.5$ & $3.5<y<4$ & $4<y<4.5$\\ \hline

0 $-$ 1&  $-30.2\pm0.7$  &  $-21.7\pm0.1$  &  $~-13.4\pm0.2$  &  $~-16.1\pm0.1$  &  $~-23.6\pm0.7$   \\  
1 $-$ 2&  $-29.1\pm0.4$  &  $-21.8\pm0.1$  &  $~-13.9\pm0.2$  &  $~-14.9\pm0.1$  &  $~-22.6\pm0.5$    \\  
2 $-$ 3&  $-29.1\pm0.3$  &  $-21.4\pm0.1$  &  $~-14.4\pm0.1$  &  $~-13.6\pm0.1$  &  $~-19.8\pm0.5$    \\  
3 $-$ 4&  $-28.3\pm0.3$  &  $-21.0\pm0.1$  &  $~-13.7\pm0.1$  &  $~-11.4\pm0.2$  &  $~-15.1\pm0.5$    \\  
4 $-$ 5&  $-27.0\pm0.3$  &  $-20.4\pm0.1$  &  $~-12.7\pm0.1$  &  $\phz~-9.3\pm0.3$  &  $~-10.6\pm0.7$    \\  
5 $-$ 6&  $-25.9\pm0.4$  &  $-20.1\pm0.1$  &  $~-10.7\pm0.2$  &  $\phz~-6.0\pm0.4$  &  $\phz~-5.6\pm0.9$    \\  
6 $-$ 7&  $-24.8\pm0.5$  &  $-17.5\pm0.1$  &  $\phz~-8.5\pm0.2$  &  $\phz~-2.9\pm0.5$  &  $\phz~-2.4\pm1.2$    \\  
7 $-$ 8&  $-23.6\pm0.5$  &  $-16.9\pm0.1$  &  $\phz~-6.7\pm0.3$  &  \multirow{2}{*}{$~~~~~\phz0.7\pm0.5$}  &  \multirow{2}{*}{$~~~~~\phz3.9\pm1.2$}    \\  
8 $-$ 9&  $-21.8\pm0.6$  &  $-16.1\pm0.2$  &  $\phz~-5.8\pm0.3$  &  $ ~ $&  $ ~ $ \\  
9 $-$ 10&  $-21.1\pm0.7$  &  $-14.7\pm0.2$  &  $\phz~-5.0\pm0.4$  &  \multirow{2}{*}{$~~~~~\phz2.7\pm0.6$}  &  \multirow{2}{*}{$~~~~\phz10.5\pm1.5$}    \\  
10 $-$ 13&  $-18.3\pm0.5$  &  $-14.1\pm0.1$  &  $\phz~-4.2\pm0.3$  &  $ ~ $&  $ ~ $  \\  
13 $-$ 20&  $-17.0\pm0.6$  &  $-11.6\pm0.2$  &  $\phz~-4.2\pm0.4$  &  $~~~~~\phz1.9\pm0.9$  &  $~~~~~\phz9.5\pm2.8$    \\  

\hline
\end{tabular}

\end{table}

\begin{table}[h]
	\centering
	\caption{Relative changes of cross-sections of $\Upsilonres(2S)$ (in \%), for a polarisation of $\lambda_{\theta}=-1$ rather than zero, in (\pt,$y$) intervals.}
        \label{table_-1pol_2S}
	\begin{tabular}{l|lllll}
\hline \pt($\!\gevc$) & $2<y<2.5$ & $2.5<y<3$ & $3<y<3.5$ & $3.5<y<4$ & $4<y<4.5$\\ \hline

0 $-$ 1&  $-29.7\pm0.7$  &  $-22.3\pm0.2$  &  $~-13.6\pm0.1$  &  $~-16.6\pm0.3$  &  $~-24.8\pm0.8$   \\  
1 $-$ 2&  $-29.5\pm0.4$  &  $-21.8\pm0.1$  &  $~-13.9\pm0.1$  &  $~-15.4\pm0.2$  &  $~-22.9\pm0.5$    \\  
2 $-$ 3&  $-29.2\pm0.3$  &  $-21.3\pm0.1$  &  $~-14.0\pm0.1$  &  $~-13.9\pm0.2$  &  $~-20.4\pm0.5$    \\  
3 $-$ 4&  $-28.2\pm0.3$  &  $-21.0\pm0.1$  &  $~-13.5\pm0.1$  &  $~-12.0\pm0.2$  &  $~-15.8\pm0.5$    \\  
4 $-$ 5&  $-27.1\pm0.3$  &  $-20.3\pm0.1$  &  $~-12.7\pm0.1$  &  $\phz~-9.4\pm0.3$  &  $~-11.4\pm0.7$    \\  
5 $-$ 6&  $-26.1\pm0.4$  &  $-20.2\pm0.1$  &  $~-11.3\pm0.1$  &  $\phz~-6.8\pm0.3$  &  $\phz~-7.3\pm0.9$   \\  
6 $-$ 7&  $-24.9\pm0.4$  &  $-18.1\pm0.1$  &  $\phz~-8.8\pm0.2$  &  $\phz~-3.5\pm0.5$  &  $\phz~-3.2\pm1.1$    \\  
7 $-$ 8&  $-24.2\pm0.5$  &  $-17.8\pm0.1$  &  $\phz~-7.1\pm0.2$  &  \multirow{2}{*}{$\phz~-0.3\pm0.5$}  &  \multirow{2}{*}{$~~~~\phz~2.7\pm1.3$}    \\  
8 $-$ 9&  $-22.2\pm0.5$  &  $-15.9\pm0.2$  &  $\phz~-7.1\pm0.2$  &  $ ~ $&  $ ~ $ \\  
9 $-$ 10&  $-22.2\pm0.6$  &  $-15.0\pm0.2$  &  $\phz~-4.5\pm0.4$  &  \multirow{2}{*}{$~~~~\phz~2.9\pm0.6$}  &  \multirow{2}{*}{$~~~~\phz~7.8\pm1.6$}   \\  
10 $-$ 13&  $-19.6\pm0.4$  &  $-14.6\pm0.2$  &  $\phz~-4.1\pm0.3$  &  $ ~ $&  $ ~ $  \\  
13 $-$ 20&  $-16.9\pm0.4$  &  $-12.1\pm0.3$  &  $\phz~-4.2\pm0.3$  &  $~~~~\phz~2.4\pm0.9$  &  $~~~~\phz~9.8\pm3.2$    \\  

\hline
\end{tabular}

\end{table}

\begin{table}[h]
	\centering
	\caption{Relative changes of cross-sections of $\Upsilonres(3S)$ (in \%), for a polarisation of $\lambda_{\theta}=-1$ rather than zero, in (\pt,$y$) intervals.}
        \label{table_-1pol_3S}
	\begin{tabular}{l|lllll}
\hline \pt($\!\gevc$) & $2<y<2.5$ & $2.5<y<3$ & $3<y<3.5$ & $3.5<y<4$ & $4<y<4.5$\\ \hline

0 $-$ 1&  $-29.7\pm0.6$  &  $-22.2\pm0.2$  &  $~-13.7\pm0.1$  &  $~-15.6\pm0.3$  &  $~-25.8\pm0.9$    \\  
1 $-$ 2&  $-29.3\pm0.4$  &  $-22.0\pm0.1$  &  $~-14.0\pm0.1$  &  $~-15.6\pm0.2$  &  $~-22.9\pm0.5$    \\  
2 $-$ 3&  $-29.1\pm0.3$  &  $-21.7\pm0.1$  &  $~-14.1\pm0.1$  &  $~-14.4\pm0.2$  &  $~-19.9\pm0.5$    \\  
3 $-$ 4&  $-28.3\pm0.3$  &  $-21.2\pm0.1$  &  $~-13.6\pm0.1$  &  $~-12.3\pm0.2$  &  $~-16.5\pm0.5$    \\  
4 $-$ 5&  $-27.9\pm0.4$  &  $-20.6\pm0.1$  &  $~-12.8\pm0.1$  &  $~-10.0\pm0.3$  &  $~-12.2\pm0.7$    \\  
5 $-$ 6&  $-26.1\pm0.4$  &  $-20.1\pm0.1$  &  $~-11.5\pm0.1$  &  $\phz~-7.3\pm0.4$  &  $\phz~-8.0\pm0.9$    \\  
6 $-$ 7&  $-25.1\pm0.4$  &  $-18.8\pm0.1$  &  $\phz~-9.6\pm0.2$  &  $\phz~-4.3\pm0.5$  &  $\phz~-4.4\pm1.1$    \\  
7 $-$ 8&  $-24.4\pm0.5$  &  $-18.0\pm0.1$  &  $\phz~-7.6\pm0.2$  &  \multirow{2}{*}{$\phz~-1.4\pm0.5$}  &  \multirow{2}{*}{$~~~~\phz~1.0\pm1.2$}    \\  
8 $-$ 9&  $-22.8\pm0.5$  &  $-15.8\pm0.2$  &  $\phz~-6.4\pm0.2$  &  $ ~ $&  $ ~ $  \\  
9 $-$ 10&  $-20.9\pm0.6$  &  $-15.7\pm0.3$  &  $\phz~-6.1\pm0.2$  &  \multirow{2}{*}{$~~~~\phz~1.7\pm0.6$}  &  \multirow{2}{*}{$~~~~\phz~8.8\pm1.6$}    \\  
10 $-$ 13&  $-19.5\pm0.4$  &  $-13.8\pm0.2$  &  $\phz~-4.6\pm0.2$  &  $ ~ $&  $ ~ $ \\  
13 $-$ 20&  $-18.0\pm0.4$  &  $-11.8\pm0.3$  &  $\phz~-5.0\pm0.2$  &  $~~~~\phz~1.7\pm0.7$  &  $~~~~\phz~9.3\pm2.9$    \\  

\hline
\end{tabular}

\end{table}

\clearpage

\addcontentsline{toc}{section}{References}
\bibliographystyle{LHCb}
\bibliography{main,standard,LHCb-PAPER,LHCb-CONF,LHCb-DP,LHCb-TDR}

\newpage
\centerline
{\large\bf LHCb collaboration}
\begin
{flushleft}
\small
R.~Aaij$^{32}$\lhcborcid{0000-0003-0533-1952},
A.S.W.~Abdelmotteleb$^{50}$\lhcborcid{0000-0001-7905-0542},
C.~Abellan~Beteta$^{44}$,
F.~Abudin{\'e}n$^{50}$\lhcborcid{0000-0002-6737-3528},
T.~Ackernley$^{54}$\lhcborcid{0000-0002-5951-3498},
B.~Adeva$^{40}$\lhcborcid{0000-0001-9756-3712},
M.~Adinolfi$^{48}$\lhcborcid{0000-0002-1326-1264},
P.~Adlarson$^{77}$\lhcborcid{0000-0001-6280-3851},
H.~Afsharnia$^{9}$,
C.~Agapopoulou$^{13}$\lhcborcid{0000-0002-2368-0147},
C.A.~Aidala$^{78}$\lhcborcid{0000-0001-9540-4988}
% ...

{\footnotesize \it

% ...
$^{9}$Universit{\'e} Clermont Auvergne, CNRS/IN2P3, LPC, Clermont-Ferrand, France\\
% ...
$^{13}$LPNHE, Sorbonne Universit{\'e}, Paris Diderot Sorbonne Paris Cit{\'e}, CNRS/IN2P3, Paris, France\\
% ...
$^{32}$Nikhef National Institute for Subatomic Physics, Amsterdam, Netherlands\\
% ...
$^{40}$Instituto Galego de F{\'\i}sica de Altas Enerx{\'\i}as (IGFAE), Universidade de Santiago de Compostela, Santiago de Compostela, Spain\\
% ...
$^{44}$Physik-Institut, Universit{\"a}t Z{\"u}rich, Z{\"u}rich, Switzerland\\
% ...
$^{48}$H.H. Wills Physics Laboratory, University of Bristol, Bristol, United Kingdom\\
% ...
$^{50}$Department of Physics, University of Warwick, Coventry, United Kingdom\\
% ...
$^{53}$School of Physics and Astronomy, University of Glasgow, Glasgow, United Kingdom\\
$^{54}$Oliver Lodge Laboratory, University of Liverpool, Liverpool, United Kingdom\\
% ...
$^{62}$Syracuse University, Syracuse, NY, United States\\
% ...
$^{77}$Department of Physics and Astronomy, Uppsala University, Uppsala, Sweden, associated to $^{53}$\\
$^{78}$University of Michigan, Ann Arbor, MI, United States, associated to $^{62}$\\
\bigskip
% ...
}
\end{flushleft}

\end{document}